\documentclass[aps,prd,11pt,tightenlines,final,preprintnumbers,
               superscriptaddress,showpacs,showkeys]{revtex4}       

\usepackage[latin1]{inputenc}                    
\usepackage{graphicx}                            
\usepackage{latexsym}                            
\usepackage{amsfonts}                            
\usepackage{amssymb}                             
\usepackage{amsmath}                             
\usepackage[mathscr]{eucal}                      
\usepackage{dcolumn}                             
\usepackage{theorem}                             
\usepackage{hyperref}                            
\graphicspath{{.}{graphics/}}                    
\newcommand{\Dlr}{\overset{\leftrightarrow}{D}}

\newcommand{\msbar}{$\overline{\mbox{\rm MS}}\ $}  

\newcommand{\cO}{\mathcal {O}}

\newcommand{\bc}{\begin{center}}
\newcommand{\ec}{\end{center}}
\newcommand{\be}{\begin{equation}}
\newcommand{\ee}{\end{equation}}
\newcommand{\bea}{\begin{eqnarray}}
\newcommand{\eea}{\end{eqnarray}}

\begin{document}

\title[Distribution amplitudes from the lattice]{Moments of pseudoscalar meson distribution amplitudes from the lattice}
\author{V.M.~Braun}\affiliation{Institut f\"ur Theoretische Physik,
  Universit\"at Regensburg, 93040 Regensburg, Germany}
\author{M.~G{\"oc}keler}\affiliation{Institut f\"ur Theoretische Physik,
  Universit\"at Regensburg, 93040 Regensburg, Germany}
\author{R.~Horsley}\affiliation{School of Physics, University of Edinburgh,
  Edinburgh EH9 3JZ, UK}
\author{H.~Perlt}\affiliation{Institut f{\"u}r Theoretische Physik,
  Universit{\"a}t Leipzig, 04109 Leipzig, Germany}
\author{D.~Pleiter}\affiliation{John von Neumann-Institut f\"ur Computing NIC /
  DESY, 15738  Zeuthen, Germany}
\author{P.E.L.~Rakow}\affiliation{Theoretical Physics Division, 
  Department of Mathematical Sciences, University of Liverpool, 
  Liverpool L69 3BX, UK}
\author{G.~Schierholz}\affiliation{Deutsches Elektronen-Synchrotron DESY,
  22603 Hamburg, Germany}
\affiliation{John von Neumann-Institut f\"ur Computing NIC /
  DESY, 15738  Zeuthen, Germany}
\author{A.~Schiller}\affiliation{Institut f{\"u}r Theoretische Physik,
  Universit{\"a}t Leipzig, 04109 Leipzig, Germany}
\author{W.~Schroers}\affiliation{John von Neumann-Institut f\"ur Computing NIC /
  DESY, 15738  Zeuthen, Germany}
\author{H.~St\"uben}
\affiliation{Konrad-Zuse-Zentrum f\"ur Informationstechnik Berlin, 
             14195 Berlin, Germany}
\author{J.M.~Zanotti}\affiliation{School of Physics, University of Edinburgh,
  Edinburgh EH9 3JZ, UK}
\collaboration{QCDSF/UKQCD collaboration} \noaffiliation

\begin{abstract}
Based on lattice simulations with two flavours of dynamical, ${\cal
  O}(a)$-improved Wilson fermions we present results for the first two
moments of the distribution amplitudes of pseudoscalar mesons at
several values of the valence quark masses. 
By extrapolating our results to the physical masses of up/down and
strange quarks, we find the first two moments of the $K^+$
distribution amplitude and the second moment of the $\pi^+$
distribution amplitude.
We use nonperturbatively determined renormalisation coefficients to
obtain results in the $\overline{\rm MS}$ scheme. 
At a scale of 4~GeV$^2$ we find $a_2^\pi=0.201(114)$ for the second
Gegenbauer moment of the pion's distribution amplitude, while for the
kaon, $a_1^K=0.0453(9)(29)$ and $a_2^K=0.175(18)(47)$. 
\end{abstract}

\pacs{12.38.Gc,14.40.Aq}

\keywords{Pseudoscalar mesons, distribution amplitudes}

\maketitle

%
%
\section{Introduction}
\setcounter{equation}{0}

In recent years exclusive reactions with identified hadrons in the
final and/or initial state are attracting increasing attention
\cite{Brodsky:1989pv}.
The reason for this interest is due to the fact that they are
dominated by rare configurations of the hadrons' constituents: either
only valence-quark configurations contribute and all quarks have small
transverse separation (hard mechanism)
\cite{Chernyak:1977as,Chernyak:1980dj,Efremov:1979qk,Efremov:1978rn,Lepage:1979zb,Lepage:1980fj,Chernyak:1977fk,Chernyak:1980dk},
or one of the partons carries most of the hadron momentum (soft or
Feynman mechanism).
In both cases, the information about hadron structure is new and
complementary to that in usual inclusive reactions, the prominent
example being the deep-inelastic lepton hadron scattering.
 
Hard contributions are simpler to treat than their soft counterparts
and their structure is well understood, see e.g.
Ref.~\cite{Stefanis:2000vd} for a recent discussion.
They can be calculated in terms of the hadron distribution amplitudes
(DAs) which describe the momentum-fraction distribution of partons at
zero transverse separation in a particular Fock state, with a fixed
number of constituents. 
DAs are ordered by increasing twist; the leading twist-2 meson DA,
$\phi_{\Pi}$, which describes the momentum distribution of the
valence quarks in the meson $\Pi$, is related to the meson's
Bethe--Salpeter wave function $\phi_{\Pi,BS}$ by an integral over
transverse momenta:
$$
\phi_{\Pi}(x,\mu^2) = Z_2(\mu^2) \int^{|k_\perp| < \mu} \!\!d^2 k_\perp\,
\phi_{\Pi,BS}(x,k_\perp).
$$
Here $x$ is the quark momentum fraction, $Z_2$ is the
renormalisation factor (in the light-cone gauge) for the quark-field
operators in the wave function, and $\mu$ denotes the renormalisation
scale.  
In particular the leading-twist DA of the pion and of the nucleon have
attracted much attention in the literature.
Furthermore, $SU(3)$ flavour symmetry breaking effects in the DAs of
strange mesons are important for predictions of the exclusive B-decay
rates (e.g.  $B\to K,K^*$) in the framework of QCD factorisation
\cite{Beneke:2000ry}, perturbative QCD \cite{Keum:2000ms},
soft-collinear effective theory (SCET)
\cite{Bauer:2000yr,Bauer:2001yt} or light-cone sum rules, e.g.
\cite{Ball:1998kk,Khodjamirian:2000ds,Ball:2004rg}.
In some cases, for instance weak radiative decays, $B\to\rho\gamma$
vs.\ $B\to K^*\gamma$, the uncertainty in $SU(3)$ breaking is actually
the dominant source of theoretical error.

The theoretical description of DAs is based on their representation
\cite{Chernyak:1977as,Chernyak:1980dj,Efremov:1979qk,Efremov:1978rn,Lepage:1979zb,Lepage:1980fj,Chernyak:1977fk,Chernyak:1980dk}
as matrix elements of a suitable nonlocal light-cone operator.
For example, for positively charged pions or kaons one defines
\begin{equation}
 \langle 0| \bar q(-z)\gamma_\mu\gamma_5 [-z,z] u(z) |\Pi^+(p)\rangle = if_\Pi p_\mu 
 \int_{-1}^1 d\xi\, e^{-i\xi p\cdot z}\phi_{\Pi}(\xi,\mu^2)\,,
\label{def:phi}
\end{equation}
where $q=d,s$, $z_\mu$ is a light-like vector, $z^2=0$, $[-z,z]$ is
the straight-line-ordered Wilson line connecting the quark and the
antiquark fields and $f_\Pi$ is the usual decay constant
$f_\pi=132$~MeV, $f_K=160$~MeV \cite{Eidelman:2004wy}.
The physical interpretation of the variable $\xi$ is that $x=
(1+\xi)/2$ and $1-x = (1-\xi)/2$ are the fractions of the meson
momentum carried by the quark and antiquark, respectively.
The definition in (\ref{def:phi}) implies the normalisation
\begin{equation}
   \int_{-1}^1 d\xi\, \phi_{\Pi}(\xi,\mu^2) = 1\,.
\end{equation}
%
For brevity, below we often drop the subscript and write $\phi$
instead of $\phi_\Pi$ unless we are referring to a specific meson.

A convenient tool to study DAs is provided by the conformal expansion
\cite{Brodsky:1980ny,Ohrndorf:1981qv,Braun:2003rp,Makeenko:1980bh}.
The underlying idea is similar to the partial-wave decomposition in
quantum mechanics and allows one to separate transverse and
longitudinal variables in the Bethe--Salpeter wave--function.  
The dependence on transverse coordinates is formulated as a scale
dependence of the relevant operators and is governed by
renormalisation-group equations. 
The dependence on the longitudinal momentum fractions is described in
terms of Gegenbauer polynomials $C^{3/2}_n(\xi)$ which are nothing but
irreducible representations of the corresponding symmetry group, the
collinear conformal group SL(2,$\mathbb R$).

In this way one obtains
\cite{Makeenko:1980bh,Brodsky:1980ny,Ohrndorf:1981qv,Efremov:1979qk,Efremov:1978rn,Lepage:1979zb,Lepage:1980fj}
\begin{equation}
\phi_\Pi(\xi,\mu^2) = \frac{3}{4} (1-\xi^2) \left(1 + \sum\limits_{n=1}^\infty
  a^\Pi_{n}(\mu^2) C_{n}^{3/2}(\xi)\right).
\label{eq:gegen1}
\end{equation}
To leading-logarithmic accuracy (LO), the (non-perturbative)
Gegenbauer moments $a_n$ renormalise multiplicatively with
\begin{equation}
a_n(\mu^2) = L^{\gamma^{(0)}_n/(2\beta_0)}\, a_n(\mu_0^2),
\end{equation}
where $L\equiv \alpha_s(\mu^2)/\alpha_s(\mu_0^2)$,
$\beta_0=11-2N_f/3$, and the anomalous dimensions $\gamma^{(0)}_n$ are
given by
\begin{equation}
\gamma^{(0)}_n =  8C_F \left( \sum_{k=1}^{n+1} \frac{1}{k} - \frac{3}{4} -
  \frac{1}{2(n+1)(n+2)} \right)
\end{equation}
with $C_F = 4/3$.
Note that the multiplicative renormalisability in leading order is not
an accident: It relies on the fact that the tree-level counterterms
retain the symmetry properties of the Lagrangian~\cite{Braun:2003rp}.

Since the anomalous dimensions increase with spin, the higher-order
contributions in the Gegenbauer expansion are suppressed at large
scales so that asymptotically only the leading term survives
\begin{equation}
 \phi(\xi,\mu^2\to\infty) = \phi_{as}(\xi) = \frac{3}{4}(1-\xi^2).
\label{def:phias}
\end{equation}
For this reason, one usually assumes that the conformal expansion is
well convergent at all scales of practical interest, and retaining the
first few terms only in the conformal expansion provides one with a
reasonable approximation for convolution integrals of the type
$\int_{-1}^1 d\xi\, \phi(\xi)/(1-\xi)$ that one encounters in many
applications.

To next-to-leading order (NLO) accuracy, the scale dependence of the
Gegenbauer moments is more complicated and reads
\cite{Mikhailov:1984ii,Mueller:1993hg,Mueller:1994cn}
\begin{equation}
 a_n(\mu^2) =  a_n(\mu_0^2) E_n^{\rm NLO}
+\frac{\alpha_s(\mu^2)}{4\pi}\sum_{k=0}^{n-2} a_k(\mu_0^2)\,  
E_k^{\rm NLO}\, d^{(1)}_{nk} \,. 
\label{eq:mom-scale}
\end{equation} 
Here we adopt the usual convention that an empty sum is equal to zero.
Moreover, $a_0=1$ and
\begin{equation}
  E_n^{\rm NLO} =  L^{\gamma^{(0)}_n/(2\beta_0)}\left[1+ 
                  \frac{\gamma^{(1)}_n \beta_0 -\gamma_n^{(0)}\beta_1}{8\pi\beta_0^2}
                 \Big[\alpha_s(\mu^2)-\alpha_s(\mu_0^2)\Big]\right] \,,
\end{equation}
where $\gamma^{(1)}_n$ are the diagonal two-loop anomalous dimensions
\cite{Mertig:1995ny}, $\beta_1 = 102-(38/3)N_f$, and the mixing
coefficients $d^{(1)}_{nk}$, $k\le n-2$, are given in closed form in
Ref.~\cite{Mueller:1993hg,Mueller:1994cn}, see also, for instance,
Ref.~\cite{Bakulev:2004cu} for a recent compilation.  
For the lowest moments $n=0,1,2$ one needs
\begin{equation}
  \gamma_0^{(1)} =0\,, \qquad \gamma_1^{(1)} = 
\frac{23488}{243} - \frac{512}{81}\, N_f\,,
   \qquad  \gamma_2^{(1)} = 
 \frac{34450}{243}-\frac{830}{81}\, N_f
\end{equation}  
and
\begin{equation}
  d^{(1)}_{20} = \frac{7}{30}( 5 C_F - \beta_0)
              \frac{\gamma^{(0)}_2}{\gamma^{(0)}_2 -2 \beta_0}
                \left[1-L^{-1+\gamma^{(0)}_2/(2\beta_0)}\right].
\end{equation}
If the mass difference between the $u$ and $d$ quarks is neglected,
$G$-parity implies that the pion DA $\phi_\pi(\xi,\mu)$ is an even
function of $\xi$, i.e all odd moments in $\xi$, $a^\pi_{2n+1}$,
vanish.  
The K-meson DA need not be even, and the calculation of $a^K_1$ will
be one of our goals.

The coefficients $a_n$ at some reference scale $\mu_0$ are
nonperturbative quantities and have to be evaluated using a
nonperturbative technique or must be extracted from experiment. 
In historic perspective, most of the discussion over the years was
centered on a particular model of the pion DA proposed by Chernyak and
Zhitnitsky in 1982 on the basis of their calculation of $a_2^\pi$
using QCD sum rules \cite{Chernyak:1981zz}. 
Using this model and assuming dominance of the hard rescattering
mechanism in exclusive reactions, Chernyak and Zhitnitsky were able to
describe an impressive amount of experimental data that were available
at that time \cite{Chernyak:1983ej}.

Since then, the original argumentation by Chernyak and Zhitnitsky and
the model itself have been largely discredited.  
Three different approaches have been used: direct calculations using
QCD sum rules, pioneered in \cite{Chernyak:1981zz}; analysis of
experimental data on the pion electromagnetic and transition form
factors (e.g. \cite{Stefanis:1998dg}) and the $B$ weak decay form
factor, using light-cone sum rules; and lattice calculations.
The summary of these results is presented in Table~2 of
Ref.~\cite{Ball:2006wn}; see also, for instance,
Refs.~\cite{Bakulev:2002uc,Bakulev:2004cu} for another recent
compilation.  
It turns out that $a_2^\pi$ can only be determined with large errors,
whatever approach is chosen.  A fair average is probably
\begin{equation}
 a_2^\pi(4\,{\rm GeV}^2) = 0.17\pm 0.15\,.
\label{a2pi}
\end{equation}

The $K$-meson DA has attracted comparatively less attention.  
The numerical value of the first moment $a^K_1$ was the subject of
significant controversy until recently.
The existing estimates are all obtained using different versions of
QCD sum rules
\cite{Khodjamirian:2004ga,Braun:2004vf,Ball:2005vx,Ball:2006fz} and
yield an average \cite{Ball:2006wn}
\begin{equation}
a_1^K(4\,{\rm GeV}^2) = 0.05\pm 0.03.
\label{a1K}
\end{equation}
For the second moment, the old estimate by Chernyak and Zhitnitsky was
$a_2^K/a_2^\pi = 0.59\pm 0.04$.  
Two recent sum rule calculations
\cite{Khodjamirian:2004ga,Ball:2006wn}, including radiative
corrections to the sum rules, give however $a_2^K/a_2^\pi\simeq 1$
pointing towards a very small $SU(3)$ violation in this coefficient.

Estimates of yet higher-order Gegenbauer coefficients are rather
uncertain.  The light-cone sum-rule calculations of the transition
form factor $F_{\pi\gamma\gamma^*}$ in
Refs.~\cite{Schmedding:1999ap,Bakulev:2002uc,Bakulev:2005cp,Agaev:2005rc}
suggest a negative value for $a_4^\pi$, which is consistent with the
result $a_4^\pi(1\,{\rm GeV}^2) > -0.07$ obtained in
Ref.~\cite{Ball:2005tb}.
However, this conclusion may be premature because yet higher moments
have been omitted (however, in Ref.~\cite{Bakulev:2001pa} they are
estimated to be small).
Moreover, there does not seem to be any convincing method to estimate
the uncertainty due to the model dependence of the analysis.
While it seems that distribution amplitude moments beyond the second
are extremely difficult to access on the lattice, it might be possible
to estimate them using the transverse lattice approach
\cite{Dalley:2002nj} or from the operator product expansion in lattice
QCD \cite{Capitani:1999fm,Detmold:2005gg}.

Last but not least, we have to mention the estimate of the pion DA in
the middle point where the momentum is shared equally between the
quark and the antiquark \cite{Braun:1988qv}
\begin{equation}
   \phi_\pi (\xi=0) = 0.6\pm 0.15\ ,
\label{eq:phi-mid}
\end{equation}  
and the measurement of the pion DA in diffractive dijet production by
E791 \cite{Aitala:2000hb}.  
Unfortunately, it turns out that collinear factorisation is broken for
dijet production \cite{Braun:2001ih,Braun:2002wu}, so that the
interpretation of this beautiful experiment is not straightforward,
see also \cite{Chernyak:2001ph,Chernyak:2001wk}.

The lattice computation of DAs of pseudoscalar $\pi$ and $K$ mesons
will be at the focus of this paper. 
On the lattice, we cannot compute non-local matrix elements of the
form (\ref{def:phi}).
However, via the light-cone operator product expansion (OPE), moments
$\langle \xi^n \rangle$ of the DAs defined by
\begin{equation}
 \langle \xi^n\rangle_\Pi (\mu^2) = \int_{-1}^1d\xi \, \xi^n \phi_\Pi
 (\xi,\mu^2)\ ,
\end{equation} 
are related to matrix elements of the local operators 
\begin{equation}
\mathcal{O}^M_{\mu_0\ldots\mu_n}(0) = i^n \bar q(0) \gamma_{\mu_0}\gamma_5 
\stackrel{\leftrightarrow}{D}_{\mu_1}\ldots 
   \stackrel{\leftrightarrow}{D}_{\mu_n} u(0)\ ,
\label{eq:op-def}
\end{equation} 
by
\begin{equation}
\langle 0| \mathcal{O}^M_{\{\mu_0\ldots\mu_n\}}(0)|\Pi(p)\rangle = 
 i f_\Pi\, p_{\{\mu_0}\ldots p_{\mu_n\}}\, \langle \xi^n \rangle_\Pi \,.
\label{eq:ope-local}
\end{equation} 
Here $M$ refers to the fact that the operator is defined in Minkowski
space, $D_\mu$ is the covariant derivative,
$\stackrel{\leftrightarrow}{D} = \stackrel{\rightarrow}{D}\! -\!
\stackrel{\leftarrow}{D}$ and $\{\ldots\}$ denotes the symmetrisation
of all indices and the subtraction of traces.
The moments $\langle \xi^n\rangle$ are related to the Gegenbauer
moments $a_n$ by simple algebraic relations:
\begin{equation}
   a_1 = \frac{5}{3}\langle \xi \rangle\,,
\qquad  a_2 = \frac{7}{12}\left(5\langle \xi^2\rangle- 1
\right)\,,\qquad \mathrm{etc.}
\label{eq:geg-alg}
\end{equation}

Although the first lattice calculation of $\langle \xi^2\rangle_\pi$
appeared almost 20 years ago \cite{Kronfeld:1984zv,Martinelli:1987si},
there has been surprisingly little activity in this area in recent
times \cite{Daniel:1990ah,DelDebbio:1999mq,DelDebbio:2002mq} to
complement other theoretical investigations.
Our preliminary result for $\langle \xi^2\rangle_\pi$ was presented in
\cite{Gockeler:2005jz} and we found in the $\overline{\rm MS}$ scheme
at $\mu^2=5\,{\rm GeV}^2$, $\langle\xi^2\rangle_\pi^{\overline{\rm
    MS}}(\mu^2=5\,{\rm GeV}^2)=0.281(28)$.
This represents the most recent lattice result. The authors of
Ref.~\cite{DelDebbio:2002mq}, on the other hand, performed a
simulation in quenched QCD and renormalised their results
perturbatively to the $\overline{\rm MS}$ scheme at
$\mu^2=7.1289$~GeV$^2$, $\langle\xi^2\rangle^{\overline{\rm
    MS}}_\pi(\mu^2=7.1289\,{\rm GeV}^2) = 0.280(49)^{+0.030}_{-0.013}\ .$

This paper is organised as follows.  
In Sec.~\ref{sec:lattice-techniques} we describe our lattice
technology including the operators and renormalisation prescriptions
used.
Section~\ref{sec:numerical-results} contains our numerical results
together with appropriate extrapolations towards the physical limits.
Finally, in Sec.~\ref{sec:summary-conclusions} we summarise our
findings by considering the results in terms of Gegenbauer moments.
Here we also discuss the insights that we gain on the shape of the
Pion and Kaon distribution amplitudes.
We tabulate our results in Appendix~\ref{sec:latt-results-work}.
%

%
%

\section{Lattice techniques}
\label{sec:lattice-techniques}
We define a meson two-point correlation function in
Euclidean space as
\begin{eqnarray}
  \lefteqn{C^{\cal O}(t,\vec{p}) = \sum_{\vec{x}} e^{-i\vec{p}\cdot\vec{x}}
    \left\langle {\cal O}_{\lbrace\mu_0\dots\mu_n\rbrace}(\vec{x},t)
      J(\vec{0},0)^\dagger \right\rangle\ ,}
  \nonumber\\
  &\rightarrow& \frac{A}{2E} \langle 0|{\cal
    O}_{\lbrace\mu_0\dots\mu_n\rbrace}(0) |\Pi(p)\rangle
  \left[e^{-Et} + \tau_{\cal O}\tau_J e^{-E(L_t-t)}\right],\ \ 
0\ll t\ll L_t\, ,
\label{eq:PS-2pt}
\end{eqnarray}
where ${\cal O}_{\lbrace\mu_0\dots\mu_n\rbrace}$ is the Euclidean
lattice transcription of Eq.~(\ref{eq:op-def}), 
\begin{equation}
  \label{eq:Eop-def}
  {\cal O}_{\lbrace\mu_0\dots\mu_n\rbrace} = \bar{q}
  \gamma_{\lbrace\mu_0} \gamma_5 \Dlr_{\mu_1} \dots
  \Dlr_{\mu_n\rbrace} u\,,
\end{equation}
$A=\langle \Pi(p)| J(0)^\dagger |0\rangle$, $E=\sqrt{m_\Pi^2 +
  \vec{p}\,^2}$, $L_t$ is the temporal extent of the lattice and we use
$J(x)\equiv\Pi(x) = \overline{q}(x)\gamma_5 u(x)$ or $J(x)\equiv
A_4(x) \equiv {\cal O}_4 = \overline{q}(x)\gamma_4\gamma_5 u(x)$ as
the interpolating operator for the pseudoscalar mesons.
The $\tau$ factor tells us how the operator behaves under time
reversal, $t\to L_t -t$.
We find that for $\tau_J$, $\tau_\Pi=-$ while $\tau_{A_4}=+$.

To increase the overlap of our interpolating operators with the ground
state, we perform Jacobi smearing at the source \cite{Best:1997qp},
while the operators we use at the sink are local.
Finally, we note that when working with operators involving
derivatives, we perform the Fourier transform in Eq.~(\ref{eq:PS-2pt})
at the ``centre-of-mass'' of the operator \cite{Gockeler:2006nb}.

\subsection{Choice of operators}
\label{sec:choice-operators}
We need to choose the lattice operators to perform the matching of the
appropriate representations of the H$(4)$ group --- the group of
Euclidean lattices relevant for our numerical computations --- to the
corresponding representations of the O$(4)$ group --- the group of
rotations and reflections in four Euclidean dimensions.

For the first moment of pseudoscalar mesons containing non-degenerate
mass quarks we consider two types of operators which we denote generically 
by ${\cal O}^a_{\mu\nu}(\mu \neq \nu)$ and ${\cal O}^b_{\mu\mu}$, e.g.
\begin{eqnarray}
{\cal O}^{a}_{41}\!\!&=&\!\! {\cal O}_{\{41\}} \, ,
  \label{eq:Oa41}\\
{\cal O}^{b}_{44}\!\!&=&\!\!{\cal
    O}_{\{44\}} - \frac{1}{3}\bigg({\cal O}_{\{11\}} + {\cal
    O}_{\{22\}} + {\cal O}_{\{33\}}\bigg)\, .
\label{eq:Ob44}
\end{eqnarray}
The first operator, ${\cal O}^{a}_{41}$, requires a nonzero momentum component
in the $1$-direction which we choose as small as possible, i.e., we
take  $\vec{p} = (p,0,0)$, where $p=2\pi/L_s$ and $L_s$ is
the spatial extent of our lattice.
Using rotational symmetry, we average over the momentum choices
$\vec{p} = (0,p,0)$ and $\vec{p} = (0,0,p)$, using the operators in
Eq.~(\ref{eq:Oa41}) with $\{41\}$ replaced with $\{42\}$ and $\{43\}$,
respectively.
The second operator, ${\cal O}^{b}_{\mu\mu}$, 
can be evaluated at $\vec{p} = (0,0,0)$.

In this situation, there will be no mixing with operators of equal or
lower dimensions, however there are improvement terms that could be
included \cite{Capitani:2000xi}.
Unfortunately the improvement coefficients are not known, so we are
forced to neglect their contribution, however they are expected to be
small and as such are unlikely to affect our results.

For the case of the second moment, which appears for mesons with both
degenerate and non-degenerate mass quarks, we also have two classes of
operators ${\cal O}^a_{\mu\nu\rho}$, ${\cal O}^b_{\mu\nu\nu}$
\cite{Gockeler:2004xb}, e.g.
\begin{eqnarray}
  {\cal O}^{a}_{412} &=& {\cal O}_{\{412\}} \, ,
  \label{eq:Oa412}\\
  {\cal O}^{b}_{411} &=& {\cal
    O}_{\{411\}} - \frac{{\cal O}_{\{422\}} + {\cal O}_{\{433\}}}{2}
\, .
\end{eqnarray}
From Eq.~(\ref{eq:ope-local}), we see that ${\cal O}^a_{\mu\nu\rho}$
requires two non-vanishing spatial components of momentum, $\vec{p} =
(p,p,0)$, while ${\cal O}^b_{\mu\nu\nu}$ needs only one, $\vec{p} =
(p,0,0)$\footnote{Here we also use rotational symmetry to average over
  the momentum choices $\vec{p} = (p,0,p)$ and $\vec{p} = (0,p,p)$,
  using the operators in Eq.~(\ref{eq:Oa412}) with $\{412\}$ replaced
  with $\{413\}$ and $\{423\}$, respectively.}.
Consideration of this fact alone would lead one to choose ${\cal
  O}^b_{\mu\nu\nu}$, since momentum components in different directions on
the lattice lead to a poorer signal.
However, lattice operators with two or more covariant derivatives can
mix with operators of the same or lower dimension. 
It turns out that for forward matrix elements, 
${\cal O}^b_{\mu\nu\nu}$ suffers from
such mixings while ${\cal O}^a_{\mu\nu\rho}$ does not.

For matrix elements involving a momentum transfer between the two
states, i.e., nonforward matrix elements, both operators ${\cal
  O}^a_{\mu\nu\rho}$ and ${\cal O}^b_{\mu\nu\nu}$ can mix with
operators involving external ordinary derivatives, i.e. operators of
the form $\partial_\mu\partial_\nu \cdots (\bar{q}\cdots q)$.
For example, ${\cal O}^{a}_{412}$ in Eq.~(\ref{eq:Oa412}) mixes only
with the following operator \cite{Gockeler:2004xb}
\begin{equation}
  {\cal O}_{412}^{a,\,\partial\partial} =
  \partial_{\{4} \partial_1 \left(\bar{q} \gamma_{2\}} \gamma_5
        q\right)\, .
\label{eq:Oa412pp}
\end{equation}
The situation for ${\cal O}^b_{\mu\nu\nu}$ is a lot worse as it can
potentially mix with up to seven different operators
\cite{Gockeler:2004xb}.
While six of these operators may vanish in the continuum limit, there
exists a mixing operator of lower dimension, and as such its
contribution must be correctly taken into account non-perturbatively.
%
Thus ${\cal O}^{a}_{\mu\nu\rho}$ offers the best possibility to
extract a value of $\langle\xi^2\rangle$ from a lattice simulation.

\subsection{\label{sec:set-gauge-fields}Set of gauge fields}
Our gauge field configurations have been generated with two flavours of
dynamical fermions, $N_f=2$, using the Wilson gluon action and
nonperturbatively ${\cal O}(a)$ improved Wilson fermions.
For four different values $\beta=5.20$, $5.25$, $5.29$, $5.40$ and up
to four different $\kappa$ values per $\beta$ we have generated ${\cal
  O}(2000-8000)$ trajectories.  
Lattice spacings and spatial volumes vary between 0.075-0.123~fm and
(1.5-2.2~fm)$^3$, respectively.  
A summary of the parameter space spanned by our dynamical
configurations can be found in Table~\ref{tab:parameters}.  
We set the scale via the force parameter, with $r_0=0.467$~fm
\cite{Aubin:2004wf,Khan:2006de}.
For more details regarding our definitions and conventions, see
Ref.~\cite{Gockeler:2005rv}.

\begin{table}[tb]
  \begin{center}
    \caption{Lattice parameters: Gauge coupling $\beta$, sea
        quark hopping parameter $\kappa_{\rm sea}$, lattice volume,
        the force scale, $r_0$, and pion mass. The latter three are
        given in lattice units.
    \label{tab:parameters}}
    \begin{tabular}{c@{\hspace{5mm}}c@{\hspace{5mm}}c@{\hspace{5mm}}c@{\hspace{5mm}}c}
        $\beta$ & $\kappa_{\rm sea}$ & Volume & $r_0/a$ & 
        $am_{\pi}$     \\ \hline
    5.20 & 0.13420 & $16^3\times 32$ & 4.077(70) & 0.5847(12)     \\
    5.20 & 0.13500 & $16^3\times 32$ & 4.754(45) & 0.4148(13)     \\
    5.20 & 0.13550 & $16^3\times 32$ & 5.041(53) & 0.2907(15)     \\
    5.25 & 0.13460 & $16^3\times 32$ & 4.737(50) & 0.4932(10)     \\
    5.25 & 0.13520 & $16^3\times 32$ & 5.138(55) & 0.3821(13)     \\
    5.25 & 0.13575 & $24^3\times 48$ & 5.532(40) & 0.25556(55)    \\
    5.29 & 0.13400 & $16^3\times 32$ & 4.813(82) & 0.5767(11)     \\
    5.29 & 0.13500 & $16^3\times 32$ & 5.227(75) & 0.42057(92)    \\
    5.29 & 0.13550 & $24^3\times 48$ & 5.566(64) & 0.32696(64)    \\
    5.29 & 0.13590 & $24^3\times 48$ & 5.840(70) & 0.23956(71)    \\
    5.40 & 0.13500 & $24^3\times 48$ & 6.092(67) & 0.40301(43)    \\
    5.40 & 0.13560 & $24^3\times 48$ & 6.381(53) & 0.31232(67)    \\
    5.40 & 0.13610 & $24^3\times 48$ & 6.714(64) & 0.22081(72) \\ \hline
    \end{tabular}
  \end{center}
\end{table}

Correlation functions are calculated on configurations taken at a
distance of 10 trajectories using 4 different locations of the fermion
source.  We use binning to obtain an effective distance of 20
trajectories.  The size of the bins has little effect on the error,
which indicates residual auto-correlations are small.

Concerning the influence of the finite size of our lattices, our
experience with other observables suggests that it is not significant
for the ensembles considered here. However, in our simulations on
smaller lattices (not included in the present analysis) finite size
effects are to be expected, the study of which is under investigation.

\subsection{\label{sec:extr-matr-elem}Extracting the matrix elements}
We calculate the average of matrix elements computed with the
following choices of momenta
\begin{eqnarray}
  {\cal O}^{a}_{\mu\nu}:\quad
  \vec{p} &=& (p,0,0)\nonumber\\
  \vec{p} &=& (0,p,0)\nonumber\\
  \vec{p} &=& (0,0,p)\, ,\nonumber\\
  {\cal O}^{b}_{\mu\mu}:\quad
  \vec{p} &=& (0,0,0)\, ,\nonumber\\
  {\cal O}^{a}_{\mu\nu\rho}:\quad
  \vec{p} &=& (p,p,0)\nonumber\\
  \vec{p} &=& (p,0,p)\nonumber\\
  \vec{p} &=& (0,p,p)\, ,\nonumber\\
\end{eqnarray}
with the indices of the operators chosen accordingly.

\begin{figure}[tb]
  \begin{center}
    \includegraphics[width=0.9\hsize]{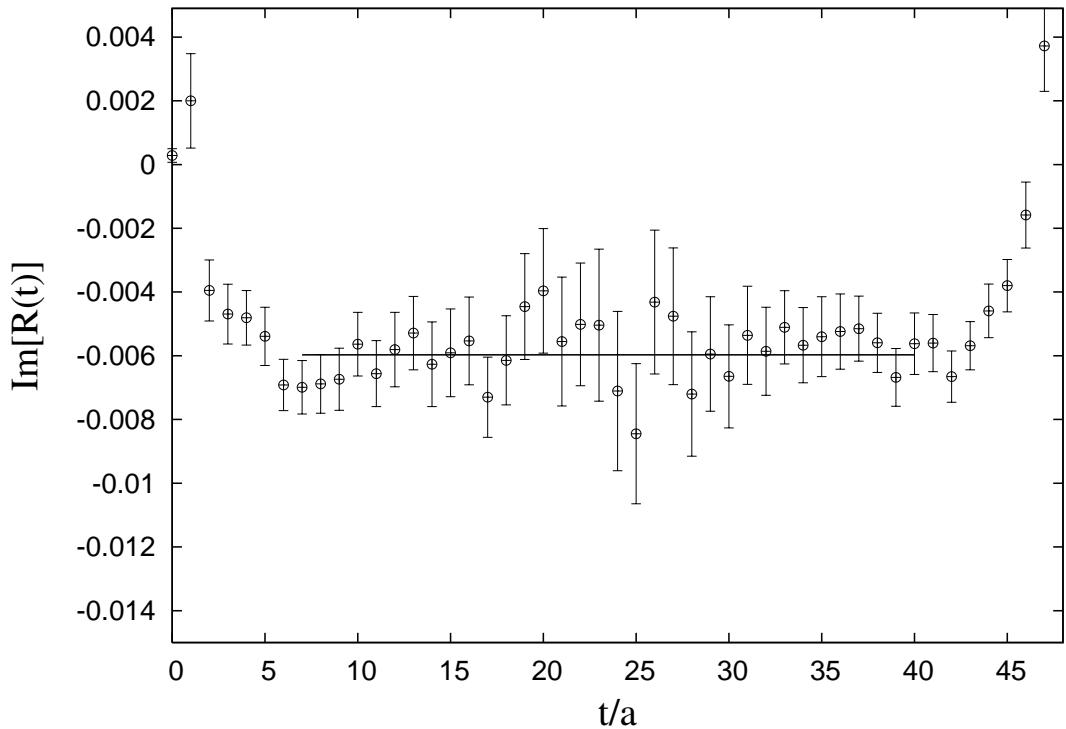}
    \caption{The imaginary part of $R^{1a}$ as defined in
      Eq.~(\ref{eq:ratio-a41}) using a $J(x)\equiv A_4(x) =
      \overline{q}(x)\gamma_4\gamma_5 u(x)$ meson interpolating field,
      for $\beta=5.29,\ \kappa_{\rm sea}=0.13550$ and valence masses,
      $\kappa_{\rm val1}=0.13550,\ \kappa_{\rm val2}=0.13430$.}
    \label{fig:ratio1a5}
  \end{center}
\end{figure}

The matrix elements of the operators given in Eqs.~(\ref{eq:Oa41}),
(\ref{eq:Ob44}), (\ref{eq:Oa412}) are then extracted from ratios of
two-point functions.
In forming the ratios from Eq.~(\ref{eq:PS-2pt}), we first need to
determine $\tau_{\cal O}$ for the various operators.
We find $\tau_{{\cal O}^a_{41}}=+,\ \tau_{{\cal O}^b_{44}}=-,\ 
\tau_{{\cal O}^a_{412}}=+$ and $\tau_{{\cal O}^b_{411}}=+$
\footnote{M.~G\"ockeler {\it et al.}, in preparation.}. 

This gives the ratios (for $0 \ll t \ll L_t$)
\begin{eqnarray}
  R^{1a} &=& \frac{C^{{\cal O}^a_{4i}}(t)}{C^{{\cal O}_4}(t)} 
    = -i\,p_i \ \langle \xi \rangle_a^\text{bare} \ ,
  \label{eq:ratio-a41} \\
  R^{1b} &=& \frac{C^{{\cal O}^b}(t)}{C^{{\cal O}_4}(t)} 
    = -\frac{E_{\vec{p}}^2 + \frac{1}{3}\vec{p}^2}{E_{\vec{p}}} \
      \langle \xi \rangle_b^\text{bare} F(E_{\vec{p}},t) ,
  \label{eq:ratio-b44} \\
  R^{2a} &=& \frac{C^{{\cal O}^a_{4ij}}(t)}{C^{{\cal O}_4}(t)} 
    = -p_i p_j \ \langle \xi^2 \rangle_a^\text{bare} \ ,
  \label{eq:ratio-a412}
\end{eqnarray}
where $i$ and $j$ are spatial indices, and ${\cal O}_4\equiv A_4(x) =
\overline{q}(x)\gamma_4\gamma_5 u(x)$ is the operator given in
Eq.~(\ref{eq:Eop-def}) with no derivatives and $\mu_0=4$.
In Eq.~(\ref{eq:ratio-b44}),
$F(E_{\vec{p}},t)=\coth{[E_{\vec{p}}(t-L_t/2)]}$ and
$\tanh{[E_{\vec{p}}(t-L_t/2)]}$ for $J(x)\equiv\Pi(x)$ and $J(x)\equiv
A_4(x)$ pseudoscalar mesons, respectively.

\begin{figure}[tb]
  \begin{center}
    \includegraphics[width=0.9\hsize]{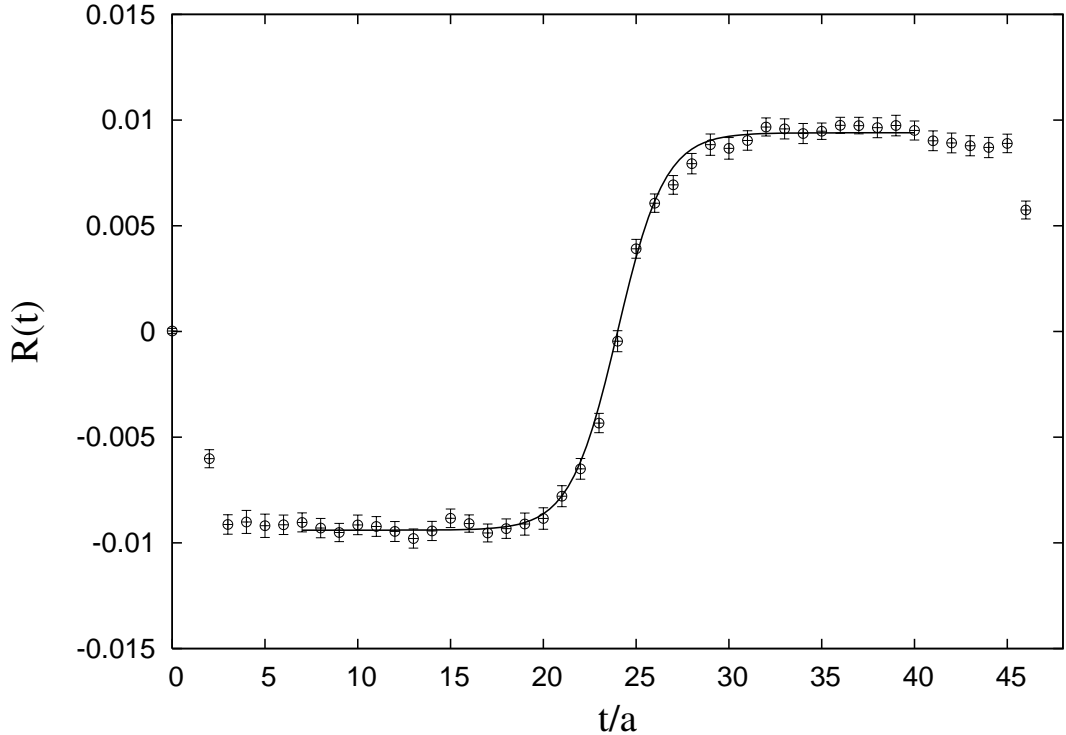}
    \caption{$R^{1b}$ as defined in Eq.~(\ref{eq:ratio-b44}) using a
      $J(x)\equiv A_4(x) = \overline{q}(x)\gamma_4\gamma_5 u(x)$ meson
      interpolating field, for $\beta=5.29,\
      \kappa_{\rm sea}=0.13550$ and valence masses, $\kappa_{\rm
        val1}=0.13550,\ \kappa_{\rm val2}=0.13430$.  Fit function is
      $y=A\tanh[b(t-L_t/2)]$, where A and $b$ are fit parameters.}
    \label{fig:ratio1b05}
  \end{center}
\end{figure}

\begin{figure}[tb]
  \begin{center}
    \includegraphics[width=0.9\hsize]{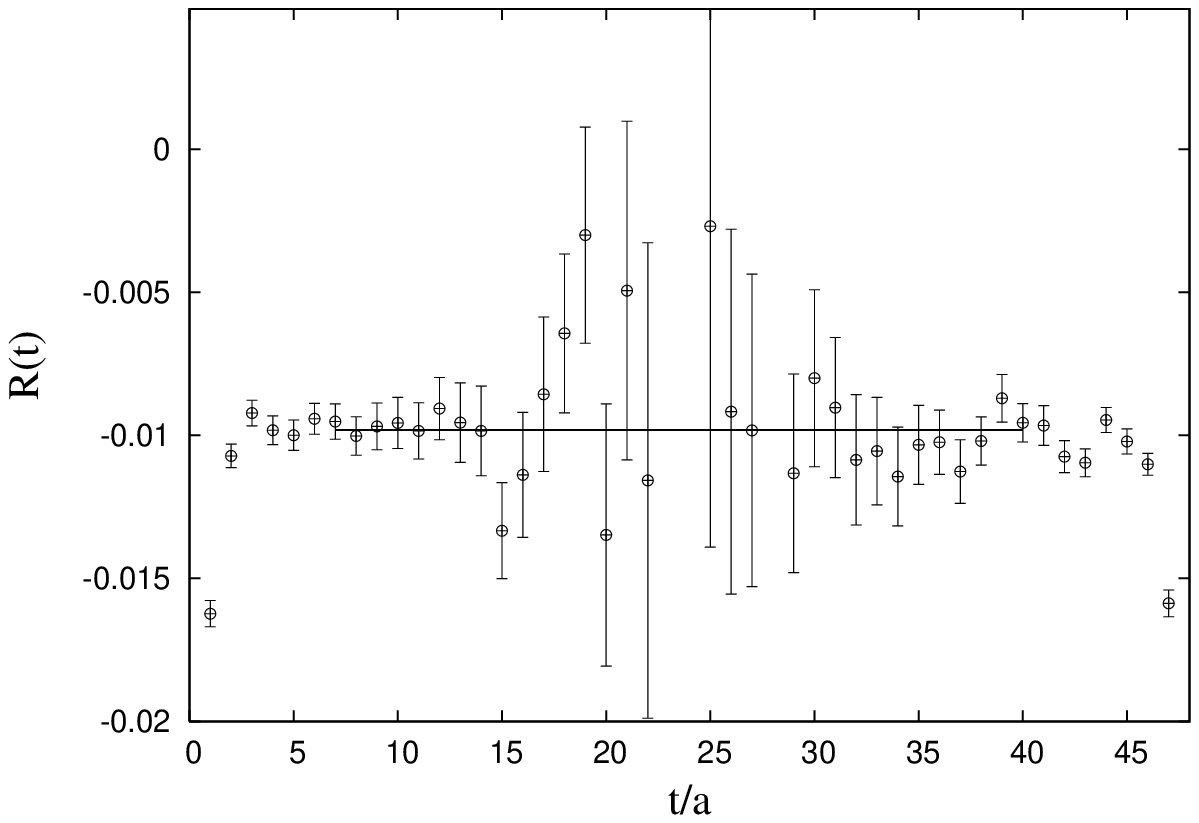}
    \caption{$R^{2a}$ as defined in Eq.~(\ref{eq:ratio-a412}) using a
      $J(x)= \overline{q}(x)\gamma_5 u(x)$ meson
      interpolating field, for $\beta=5.29,\
      \kappa_{\rm sea}=0.13590$ and degenerate valence masses,
      $\kappa_{\rm val1}=0.13490,\ \kappa_{\rm val2}=0.13490$.}
    \label{fig:ratio2a5}
  \end{center}
\end{figure}

Figure~\ref{fig:ratio1a5} shows a typical example of the ratio in
Eq.~(\ref{eq:ratio-a41}) using a $J(x)\equiv A_4(x)$ pseudoscalar
meson ($\langle\xi\rangle_a^{45}$), where we observe a plateau between
$t=7$ and $t=40$.
After extracting $R^{1a}$ from the plateaus, we use
Eq.~(\ref{eq:ratio-a41}) to extract $\langle \xi
\rangle_a^\text{bare}$.
Similarly, a hyperbolic tangent fit to the ratio $R^{1b}$ in
Fig.~\ref{fig:ratio1b05} and a constant fit to $R^{2a}$ in
Fig.~\ref{fig:ratio2a5} allow for the extraction of $\langle \xi
\rangle_b^\text{bare}$ and $\langle \xi^2 \rangle_a^\text{bare}$,
respectively.

Here and in the following, we use the notation $\langle\xi^n\rangle^5$
and $\langle\xi^n\rangle^{45}$ to distinguish the results for
$J(x)\equiv\Pi(x)$ and $J(x)\equiv A_4(x)$ pseudoscalar mesons,
respectively.

\subsection{\label{sec:operator-renorm}Operator Renormalisation and Mixing}
In general, bare lattice operators must be renormalised in some scheme
${\cal S}$ and at a scale $M$. If the operator is multiplicatively
renormalisable, which is the case for the operators (\ref{eq:Oa41}) and 
(\ref{eq:Ob44}), we have
\be
{\cal O}^{\cal S}(M^2) = Z^{\cal S}_{\cal O}(M^2) {\cal O}(a)\ ,
\ee
where ${\cal O}(a)$ denotes the bare operator at lattice spacing $a$.
Since $\langle \xi^n \rangle$ is computed from a ratio of 
two-point functions with the operator $\cal O$ in the numerator 
and the 4-component of the axial vector current ${\cal O}_4$ in the 
denominator the renormalised value is given by
\be
\langle \xi^n \rangle^{\cal S}(M^2) = \frac{Z^{\cal S}_{\cal O}(M^2)}
{Z_{{\cal O}_4}} \langle\xi^n\rangle^{\rm bare}\ ,
\label{eq:renorm-xin}
\ee
if $\cal O$ is multiplicatively renormalisable.

In this work, we renormalise our operators non-perturbatively.  
Here we restrict ourselves to a short outline of the procedure.
More details can be found in Section~5.2.3 of
Ref.~\cite{Gockeler:2004wp}, and a fuller account will be given in a
forthcoming publication.

We start from a MOM-like renormalisation condition imposed on the
lattice~\cite{Martinelli:1994ty,Gockeler:1998ye} and perform a chiral
extrapolation of the non-perturbative renormalisation factors at fixed
$\beta$ and fixed momentum.
We then apply continuum perturbation theory to calculate the
renormalisation group invariant renormalisation factor $Z^{\rm RGI}$
from the chirally extrapolated $Z$s \cite{Gockeler:2004wp}.
Our results for the operators (\ref{eq:Oa41}) and (\ref{eq:Ob44}),
i.e.\ $Z_{1a}^\text{RGI}$ and $Z_{1b}^\text{RGI}$, can be found in
Table~\ref{table:NP-Z}, where also $Z_{{\cal O}_4}$ is given.
Note that $Z$ and $Z^{\rm RGI}$ coincide for ${\cal O}_4$ because the
anomalous dimension of the axial vector current vanishes.

\begin{center}
\begin{table}[tb]
\caption{Results for the non-perturbative RGI renormalisation
  constants, $Z^\text{RGI}$, for the operators defined in
  Eqs.~(\ref{eq:Oa41}), (\ref{eq:Ob44}) and (\ref{eq:Oa412}) 
  as well as for ${\cal O}_4$.} 
\begin{tabular}{c@{\hspace{0.9cm}}cccc}
$\beta$  &  $Z_{1a}^\text{RGI}$  &  $Z_{1b}^\text{RGI}$  & $Z_{2a}^\text{RGI}$
& $Z_{{\cal O}_4}$ \\
\hline
5.20     &  1.52(4) &  1.55(5) &  2.4(1)   & 0.765(5) \\
5.25     &  1.52(4) &  1.55(5) &  2.4(1)   & 0.769(4) \\
5.29     &  1.54(4) &  1.56(5) &  2.45(10) & 0.772(4) \\
5.40     &  1.57(3) &  1.60(4) &  2.5(1)   & 0.783(4) \\
\hline
\end{tabular}
\label{table:NP-Z}
\end{table}
\end{center}

In the final step we have to convert $Z^{\rm RGI}$ to the
$\overline{\rm MS}$ scheme at some renormalisation scale $M^2=\mu^2$.
This is done perturbatively, and the result depends on the value of
$\Lambda_{\overline{\rm MS}}$ in physical units. We use
$r_0\Lambda_{\overline{\rm MS}}=0.617$ \cite{Aubin:2004wf}~%
and $r_0 = 0.467$ fm to obtain
$\Lambda_{\overline{\rm MS}} = 261$ MeV. For the operators 
(\ref{eq:Oa41}) and (\ref{eq:Ob44}) we find 
$Z^{\overline{\rm MS}}/Z^{\rm RGI} = 0.7154$ at the scale 
$\mu^2 = 4 \, \mbox{GeV}^2$.

If there are operators having the same quantum numbers and
the same or lower dimension, 
they may mix with the operator we are interested in
and we must renormalise our operator via
\be
{\cal O}^{\cal S}_i(M^2) = \sum_j
    Z_{ij}^{\cal S}(M^2, a) {\cal O}_j(a)\, .
\ee

From \cite{Gockeler:2004xb} we know that ${\cal O}^a_{412}$
(Eq.~(\ref{eq:Oa412})) mixes with ${\cal
  O}^{a,\,\partial\partial}_{412}$ (Eq.~(\ref{eq:Oa412pp}))
such that the renormalised operator can be written as
\begin{equation} 
 {\cO}^{\cal S}_{412} = 
          Z^{\cal S}_{412} {\cO}^a_{412} + Z^{\cal S}_\text{mix}
          {\cO}^{a,\partial  \partial}_{412} \,.
\label{eq:reno412}
\end{equation}
If we restrict ourselves to forward matrix elements, the operator
${\cal O}^{a,\,\partial\partial}_{412}$ cannot contribute and ${\cal
  O}^a_{412}$ becomes effectively multiplicatively renormalisable.
Thus we can compute $Z^{\cal S}_{412}$ in Eq.~(\ref{eq:reno412})
non-perturbatively as sketched above. 
A sample result is shown in Fig.~2 of Ref.~\cite{Gockeler:2005vw},
where $Z^{\mathrm {RGI}}_{2a}$ is called $Z^{\{5\}}_{\mathrm {RGI}}$.
Our numbers for $Z_{2a}^\text{RGI}$ are also given in
Table~\ref{table:NP-Z}.

The mixing factor $Z^{\cal S}_\text{mix}$, on the other hand, has only
been computed in one-loop tadpole-improved lattice perturbation
theory~\cite{Gockeler:2006nb}.
In order to avoid the logarithms in the perturbative expressions
we work at the scale $\mu^2 = 1/a^2$, where $a$ is obtained from the 
value of $r_0/a$ in the chiral limit \cite{Gockeler:2005rv}. In this way
we find the numbers given in Table~\ref{table:Z1}. 

\begin{center}
\begin{table}
\caption{Results for the renormalisation mixing
  coefficient, $Z^{\overline{\rm MS}}_\text{mix}$, computed in
  tadpole-improved perturbation theory in the $\overline{\rm MS}$
  scheme at the scale $\mu^2 = (1/a)^2$,
  where $a$ is obtained from the value of $r_0/a$ in the chiral limit
  \cite{Gockeler:2005rv}.} 
\begin{tabular}{c@{\hspace{0.9cm}}c@{\hspace{0.9cm}}c}
$\beta$  &  $\mu^2=1/a^2$ [GeV$^2$] &  
$Z^{\overline{\rm MS}}_\text{mix}$   \\
\hline
5.20   &   5.3361   &  $-0.00258$ \\
5.25   &   6.2001   &  $-0.00253$ \\
5.29   &   6.9696   &  $-0.00250$ \\
5.40   &   9.7344   &  $-0.00240$ \\
\hline
\end{tabular}
\label{table:Z1}
\end{table}
\end{center}

The values of the conversion factor $Z^{\overline{\rm MS}}_{2a}/Z^{\rm
  RGI}_{2a}$ at the relevant scales are again computed in continuum
perturbation theory and are collected in Table~\ref{table:Zconv}.

\begin{center}
\begin{table}
\caption{Results for $Z^{\overline{\rm MS}}_{2a}/Z^{\rm RGI}_{2a}$ 
at $\mu^2 = (1/a)^2$ for the lattice spacings $a$ in our simulations.}
\begin{tabular}{c@{\hspace{0.9cm}}c@{\hspace{0.9cm}}c}
$\beta$  &  $\mu^2=(1/a)^2$ [GeV$^2$] & $Z^{\overline{\rm MS}}_{2a}/Z^{\rm RGI}_{2a}$  \\
\hline
5.20     &  5.3361  & 0.5650 \\
5.25     &  6.2001  & 0.5545 \\
5.29     &  6.9696  & 0.5465 \\
5.40     &  9.7344  & 0.5262 \\
\hline
\end{tabular}
\label{table:Zconv}
\end{table}
\end{center}

Denoting the unrenormalised values of $f_\Pi$ and
$\langle \xi^2 \rangle$ by $f^{\mathrm {bare}}_\Pi$ and 
$\langle \xi^2 \rangle^{\mathrm {bare}}$, respectively, we have
from Eq.~(\ref{eq:ope-local})
\begin{equation} 
 \langle 0 |{\cO}^a_{412} | \Pi (p) \rangle
  = f^{\mathrm {bare}}_\Pi p_1 p_2 p_4 \langle \xi^2 \rangle^{\mathrm
    {bare}}\ ,
\end{equation}
and
\begin{equation} \begin{array}{l} \displaystyle
 \langle 0 |{\cO}^{\cal S}_{412} | \Pi (p) \rangle
  = f^{\mathrm {bare}}_\Pi p_1 p_2 p_4 
 \left( Z^{\cal S}_{412} \langle \xi^2 \rangle^{\mathrm {bare}} + Z^{\cal
     S}_\text{mix} \right) 
\\  \displaystyle 
  \hphantom{\langle 0 |{\cO}^{\mathrm R}_{412} | \Pi (p) \rangle} {}
  = f_\Pi p_1 p_2 p_4  \left( \frac{Z^{\cal S}_{412}}{Z_{{\cO}_4}} 
\langle \xi^2 \rangle^{\mathrm {bare}} + \frac{Z^{\cal
    S}_\text{mix}}{Z_{{\cO}_4}}
\right) \,.
\end{array} \end{equation}
Here the renormalised $f_\Pi$ is given by
\begin{equation} 
 f_\Pi = Z_{{\cO}_4} f^{\mathrm {bare}}_\Pi \ ,
\end{equation}
and for the renormalised $\langle \xi^2 \rangle$ we get
\begin{equation} 
\langle \xi^2 \rangle = \frac{Z^{\cal S}_{412}}{Z_{{\cO}_4}}
           \langle \xi^2 \rangle^{\mathrm {bare}} +
           \frac{Z^{\cal S}_\text{mix}}{Z_{{\cO}_4}}\,.
\label{eq:renorm-xi2}
\end{equation}

So we first obtain $\langle \xi^2 \rangle$ at the scale $\mu^2_0 =
(1/a)^2$.
Using the relation between $\langle \xi^2 \rangle$ and the Gegenbauer
moment $a_2$, Eq.~(\ref{eq:geg-alg}), along with the NLO scale
dependence of the latter, Eq.~(\ref{eq:mom-scale}), we get $\langle
\xi^2 \rangle$ at the scale $\mu^2 = 4 \, \mbox{GeV}^2$.
We calculate the running coupling from the 4-loop approximation of the
$\beta$-function in the $\overline{\mathrm {MS}}$ scheme with
$\Lambda_{\overline{\mathrm {MS}}} = 0.261 \, \mbox{GeV}$
\cite{Aubin:2004wf}.

%
%

\section{\label{sec:numerical-results}Numerical results}

\subsection{\label{sec:flav-diag-meson}Mesons with mass degenerate quarks}
Investigating quark mass degenerate mesons, i.e., the matrix element
Eq.~(\ref{eq:ope-local}) using the operator in Eq.~(\ref{eq:op-def})
with identical masses for the fermion propagators, allows us to
investigate the structure of the pions. 
In this case, all odd moments vanish, hence we focus on the lowest
non-trivial moment, $\langle\xi^2\rangle$.

For each of our datasets, we extract a value for
$\langle\xi^2\rangle^{\rm bare}$ from Eq.~(\ref{eq:ratio-a412}) and
renormalise using Eq.~(\ref{eq:renorm-xi2}).
In Table~\ref{table:xi2} we present our results for
$\langle\xi^2\rangle^{\rm bare}$.
We find that the results for $\langle\xi^2\rangle$ using the $A_4$
meson interpolating operator lead to very poorly constrained chiral
and continuum extrapolations for operators involving 2 derivatives.
Hence in the following we only discuss the results for
$\langle\xi^2\rangle$ obtained using the $\Pi$ interpolating field.

\begin{figure}[t]
\bc
\includegraphics[width=0.9\hsize]{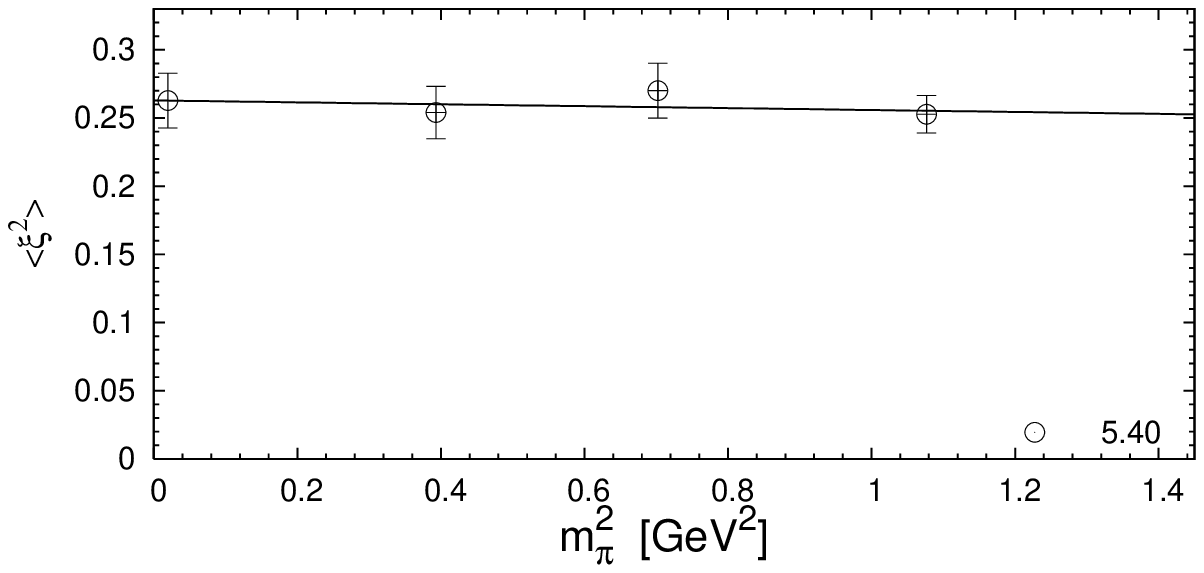}
\includegraphics[width=0.9\hsize]{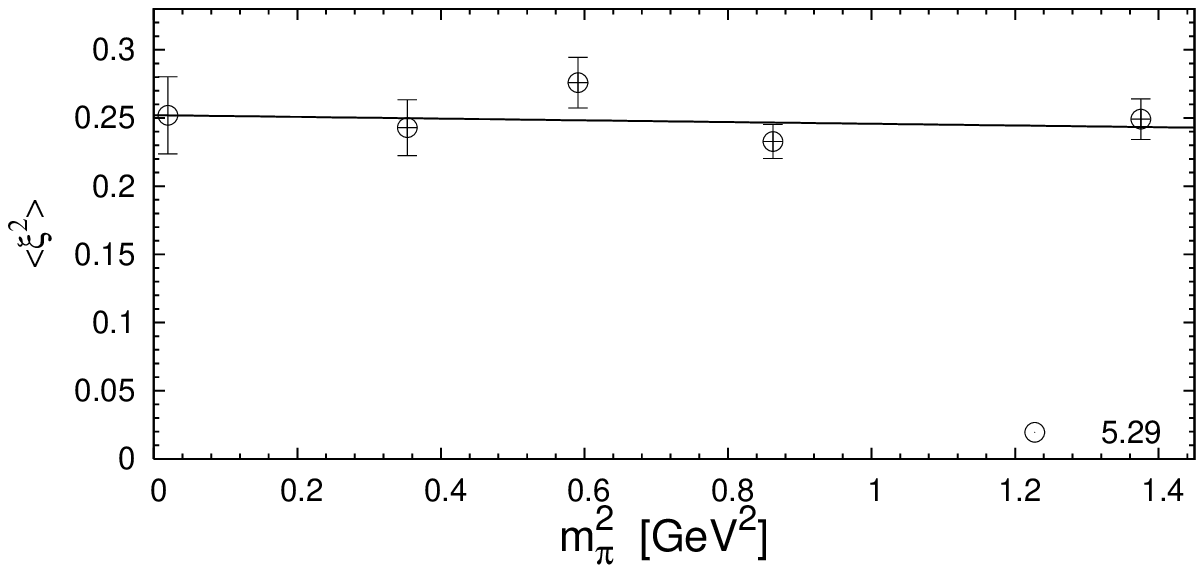}
\caption{Chiral extrapolation of $\langle\xi^2\rangle_\pi$ at constant
  $\beta$ for $\beta=5.40$ (top) and $\beta=5.29$ (bottom) for ${\cal
    O}^a_{412}$ from Eq.~(\ref{eq:Oa412}) in the $\overline{\rm MS}$
  scheme at $\mu^2=4\,\text{GeV}^2$.}
\label{fig:xi2-chiral}
\ec
\vspace*{-5mm}
\end{figure}

\begin{figure}
\bc
\includegraphics[width=0.9\hsize]{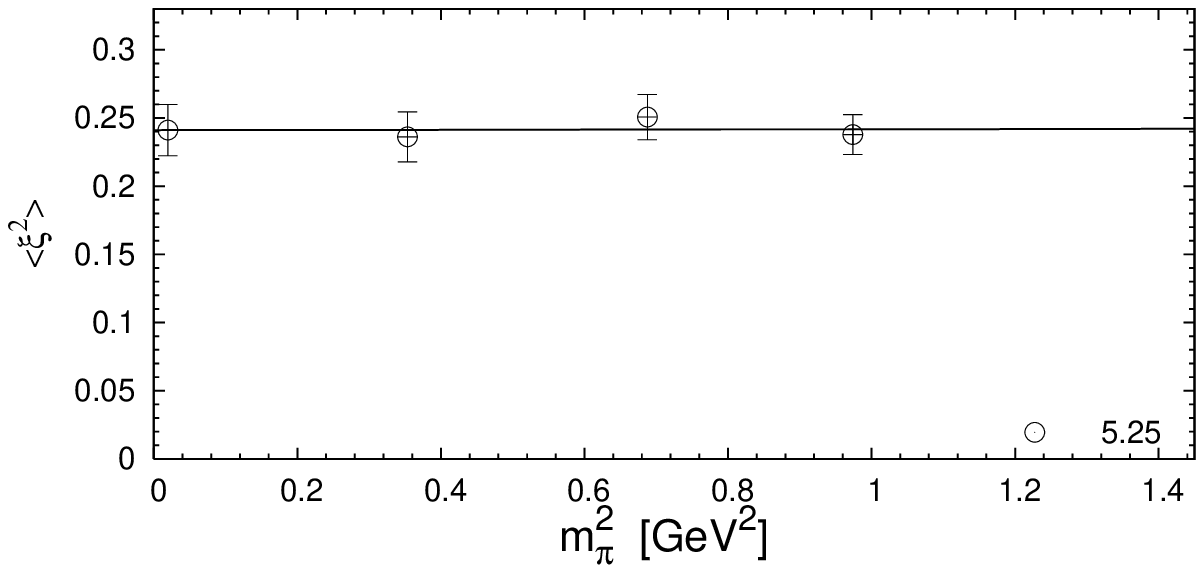}
\includegraphics[width=0.9\hsize]{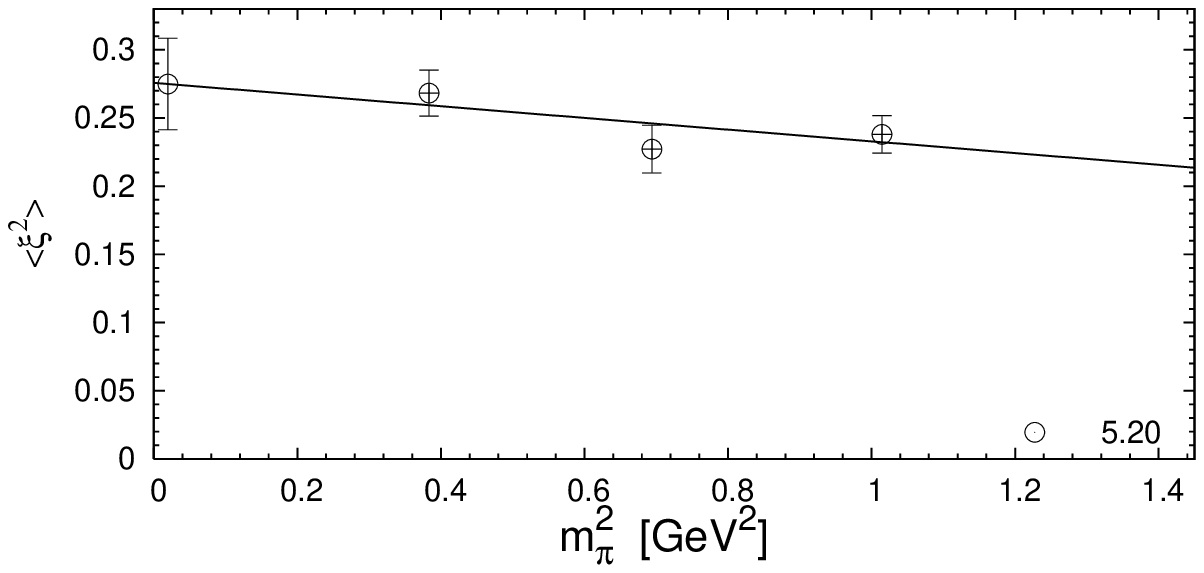}
\caption{Chiral extrapolation of $\langle\xi^2\rangle_\pi$ at constant
  $\beta$ for $\beta=5.25$ (top) and $\beta=5.20$ (bottom) for ${\cal
    O}^a_{412}$ from Eq.~(\ref{eq:Oa412}) in the $\overline{\rm MS}$
  scheme at $\mu^2=4\,\text{GeV}^2$.}
\label{fig:xi2-chiral2}
\ec
\vspace*{-5mm}
\end{figure}

In order to obtain a result in the continuum limit at the physical
pion mass, we first extrapolate our results at constant $\beta$ to the
physical pion mass.
In Fig.~\ref{fig:xi2-chiral} we display the chiral extrapolations for
$\beta=5.40$ (top) and 5.29 (bottom), while Fig.~\ref{fig:xi2-chiral2}
contains the corresponding extrapolations for $\beta=5.25$ (top) and
5.20 (bottom).
These results exhibit only a mild dependence on the quark mass and
their values at the physical pion mass agree within errors.
The smooth linear behaviour of $\langle\xi^2\rangle$ was predicted in
Ref.~\cite{Chen:2003fp,Chen:2005js} where it was shown that at
next-to-leading order in chiral perturbation theory, all possible
non-analytic corrections to the matrix elements (\ref{eq:ope-local})
are contained in $f_\Pi$.

\begin{table}[tb]
  \begin{center}
    \caption{Bare results for $\langle\xi^2\rangle^{5}_a$ and
      $\langle\xi^2\rangle^{45}_a$ calculated on each dataset with
      degenerate valence quark masses
      $\kappa_{\text{val}}=\kappa_\text{sea}$.
    \label{table:xi2}}
    \begin{tabular}{c@{\hspace{2mm}}|@{\hspace{2mm}}c@{\hspace{2mm}}|@{\hspace{2mm}}c@{\hspace{2mm}}|@{\hspace{2mm}}c}
        $\beta$ & $\kappa_{\rm sea}$ & $\langle\xi^2\rangle_a^{5}$ 
& $\langle\xi^2\rangle_a^{45}$     \\ \hline
    5.20 & 0.13420 & 0.1353(47) & 0.1447(46)    \\
    5.20 & 0.13500 & 0.1296(77) & 0.1575(62)    \\
    5.20 & 0.13550 & 0.1518(65) & 0.140(10)    \\
    5.25 & 0.13460 & 0.1380(55) & 0.1328(82)    \\
    5.25 & 0.13520 & 0.1450(67) & 0.1706(57)    \\
    5.25 & 0.13575 & 0.1371(82) & 0.1541(93)    \\
    5.29 & 0.13400 & 0.1434(54) & 0.1537(47)    \\
    5.29 & 0.13500 & 0.1346(37) & 0.1587(35)    \\
    5.29 & 0.13550 & 0.1578(76) & 0.1737(68)    \\
    5.29 & 0.13590 & 0.1401(94) & 0.1769(77)    \\
    5.40 & 0.13500 & 0.1488(42) & 0.1516(58)    \\
    5.40 & 0.13560 & 0.1581(86) & 0.1780(74)    \\
    5.40 & 0.13610 & 0.1495(83) & 0.172(11) \\ \hline
    \end{tabular}
  \end{center}
\end{table}

\begin{figure}[t]
\bc
\includegraphics[width=0.9\hsize]{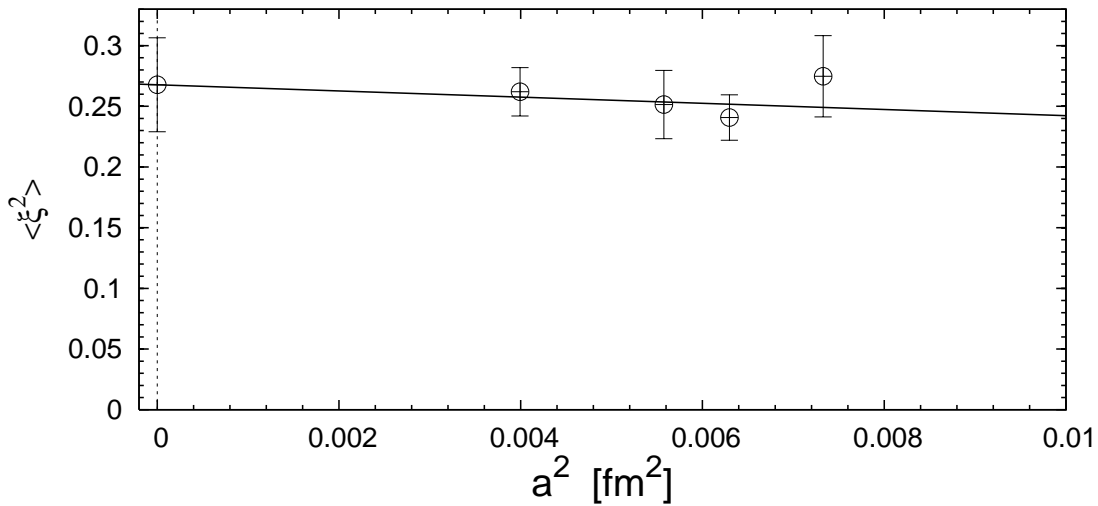}
\caption{Results for $\langle\xi^2\rangle_\pi$ for each value of
  $\beta$ at the physical pion mass as a function of $a^2$ for ${\cal
    O}^a_{412}$ from Eq.~(\ref{eq:Oa412}) in the $\overline{\rm MS}$
  scheme at $\mu^2=4\,\text{GeV}^2$.}
\label{fig:xi2-cont}
\ec
\vspace*{-5mm}
\end{figure}

Now that we have calculated results at the physical pion mass for each
choice of $\beta$, we are in a position to examine the behaviour of
our results as a function of the lattice spacing.
In Fig.~\ref{fig:xi2-cont} we use the values of $r_0/a$ extrapolated
to the chiral limit for each $\beta$ (see Table~3 of
Ref.~\cite{Gockeler:2005rv}) to study the dependence of our results on
the lattice spacing.
Even though our operators are not ${\cal O}(a)$-improved, we find a
negligible dependence on the lattice spacing, at least when compared
to the statistical errors.

Employing a linear extrapolation to the continuum limit at the
physical pion mass, we find the second moment of the pion's
distribution amplitude to be
\be 
\langle\xi^2\rangle^{\overline{\rm MS}}_\pi(\mu^2=4\,{\rm GeV}^2)
= 0.269(39)\ ,
\label{eq:xi2pi-result}
\ee
with an acceptable $\chi^2/dof=0.5$, which is close to the value
$\langle\xi^2\rangle^{\overline{\rm MS}}_\pi(\mu^2=4\,{\rm
  GeV}^2) = 0.286(49)^{+0.030}_{-0.013}$ found in
Ref.~\cite{DelDebbio:2002mq}, and larger than the asymptotic value,
$\langle\xi^2\rangle_{\mbox{\tiny as}} = 0.2$.

\begin{figure}[tb]
  \begin{center}
    \includegraphics[width=0.9\hsize]{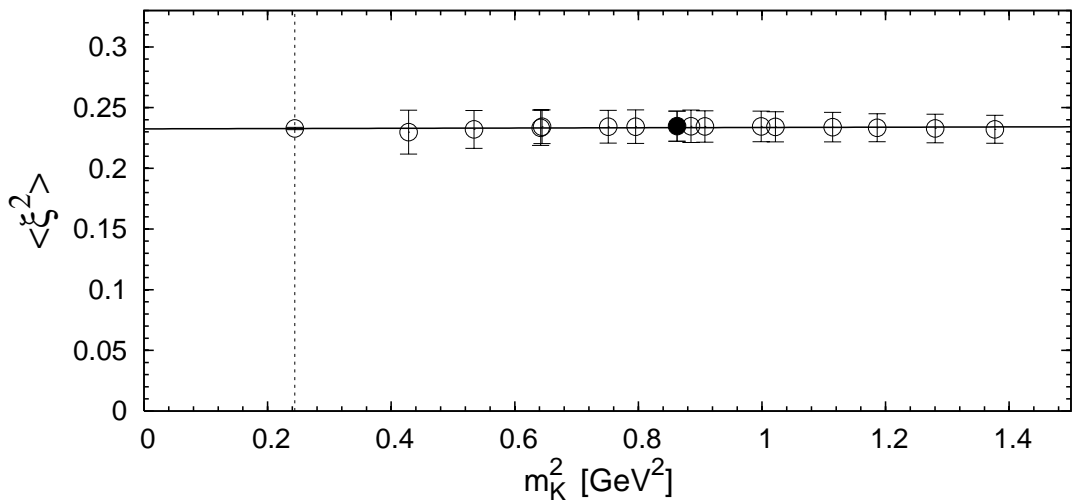}
    \caption{$\langle\xi^2\rangle_K$, extracted from Eq.~(\ref{eq:ratio-a412}) at the
      working point, $\beta=5.29$, $\kappa_{\text{sea}}=0.13500$, as a
      function of the squared Kaon mass, $m_K^2$, for various choices
      of the valence quark masses. Results are quoted in the
      $\overline{\rm MS}$ scheme at $\mu^2=4\,\text{GeV}^2$. The
      vertical dotted line corresponds to the physical Kaon mass.}
    \label{fig:b5p29kp13500_2a5}
  \end{center}
\end{figure}

\begin{figure}[tb]
  \begin{center}
    \includegraphics[width=0.9\hsize]{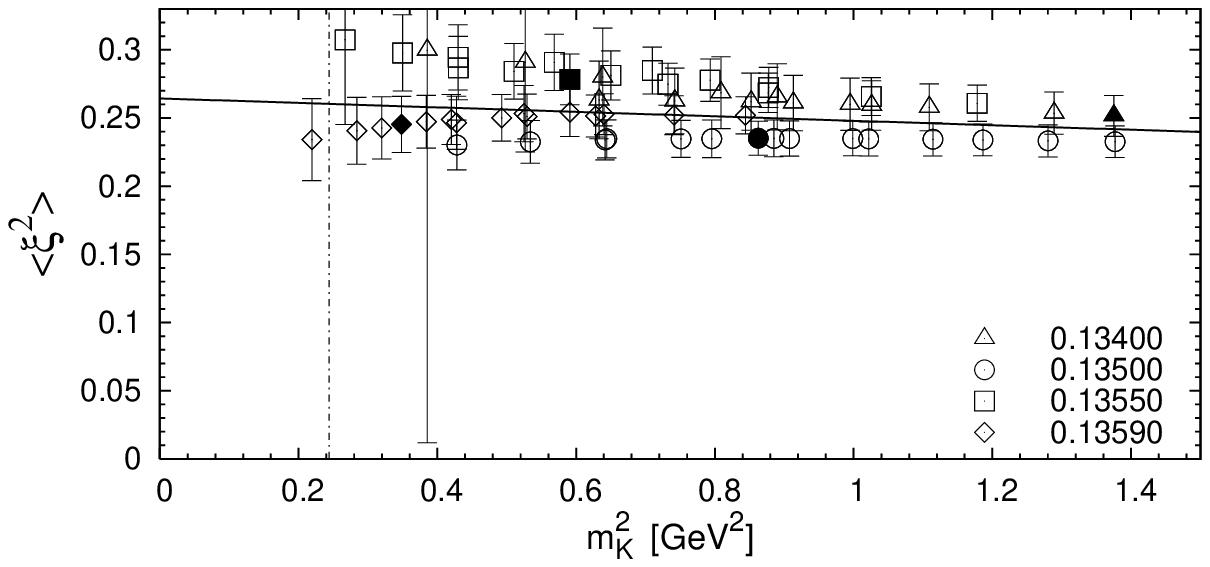}
    \caption{Results of fit in
      Eq.~(\ref{eq:xi2-fit}) for $m_\pi=m_\pi^\text{physical}$,
      together with our results for $\langle\xi^2\rangle_K$ from all
      four values of $\kappa_\text{sea}$ at $\beta=5.29$ considered.
      Results are quoted in the $\overline{\rm MS}$ scheme at
      $\mu^2=4\,\text{GeV}^2$. The vertical dotted line corresponds to
      the physical Kaon mass.}
    \label{fig:b5p29_2a5}
  \end{center}
\end{figure}

\subsection{\label{sec:flav-offdiag-meson}Mesons with mass
  non-degenerate quarks}
When the masses of the quark and the antiquark in
Eq.~(\ref{eq:ope-local}) become unequal, the odd moments will no longer
vanish and --- with appropriate adjustment of the quark masses --- we
can directly obtain the corresponding moments of the Kaon.
The results that will be discussed in this section are tabulated in
Tables~\ref{table:b5p29kp13400}, \ref{table:b5p29kp13500},
\ref{table:b5p29kp13550} and \ref{table:b5p29kp13590}.
Due to the large amount of resources required to calculate these
partially quenched results, we simulate at a fixed value of
$\beta=5.29$ where we have four different sea quark masses at our
disposal.
As a result, we are not able to examine the lattice spacing dependence of
these results.
However, we take encouragement from our results in the previous
section, where we found that discretisation effects are small for
$\langle\xi^2\rangle$, and neglect the extrapolation to the continuum limit.

Occasionally the raw data is so noisy that it is not possible to
perform a stable fit to one or more of the ratios in
Eqs.~(\ref{eq:ratio-a41}), (\ref{eq:ratio-b44}) and
(\ref{eq:ratio-a412}).
In such instances, we are unable to report a result and hence gaps are
present in Tables~\ref{table:b5p29kp13400}, \ref{table:b5p29kp13500},
\ref{table:b5p29kp13550} and \ref{table:b5p29kp13590}.

\subsubsection{Second moment}
Figure~\ref{fig:b5p29kp13500_2a5} shows the second moment,
$\langle\xi^2\rangle_K$, extracted from Eq.~(\ref{eq:ratio-a412}) at the
working point, $\beta=5.29$, $\kappa_{\text{sea}}=0.13500$, as a
function of the squared Kaon mass, $m_K^2$, for various choices of the
valence quark masses.
Here when we refer to the Kaon mass, we mean a pseudoscalar mass which
is a function of two valence quarks,
$m_K=m_\text{ps}(\kappa_\text{val1},\,\kappa_\text{val2})$, where
$\kappa_\text{val1}\ge\kappa_\text{val2}$.
(These masses are provided in the third columns of
Tables~\ref{table:b5p29kp13400}--\ref{table:b5p29kp13590}.)
A solid symbol indicates the point where
$\kappa_\text{val1}=\kappa_\text{val2}=\kappa_\text{sea}$.
The behaviour towards the chiral limit of the available data points
suggests that indeed a linear extrapolation is possible.
The vertical dotted line indicates the physical kaon mass, $m_K=0.494$~GeV.

In order to obtain a result at the physical $\pi$ and $K$ masses, we
performed similar fits at all available sea quark masses corresponding to
$\kappa_\text{sea}=0.13400, 0.13500, 0.13550, 0.13590$, and then
tried to extrapolate in the sea quark mass (or
$m_\pi=m_\text{ps}(\kappa_\text{sea},\,\kappa_\text{sea})$) to
$m_\pi=0.140$~GeV. 
This final extrapolation, however, turns out to be unreliable (large
$\chi^2/dof$). 
Hence we attempt to fit to all the data available with the global
ansatz
\be
\langle\xi^2\rangle_K = \alpha + 
\beta m_\pi^2(\kappa_\text{sea},\,\kappa_\text{sea}) +
\gamma m_K^2(\kappa_\text{val1},\,\kappa_\text{val2})\ ,
\label{eq:xi2-fit}
\ee
with three fit parameters, $\alpha,\,\beta,\,\gamma$.
After performing such a fit we find
\be
\alpha = 0.264(7),\ \beta = -0.00005(841),\ \gamma = -0.016(9)\ ,
\label{eq:xi2-fit-results}
\ee
with a $\chi^2/dof = 1.06$.
The fit results indicate that the dependence of
$\langle\xi^2\rangle_K$ on the sea quark mass is negligible, while the
dependence on the valence quarks is very small.

\begin{figure}[tbp!]
  \begin{center}
    \includegraphics[width=0.8\hsize]{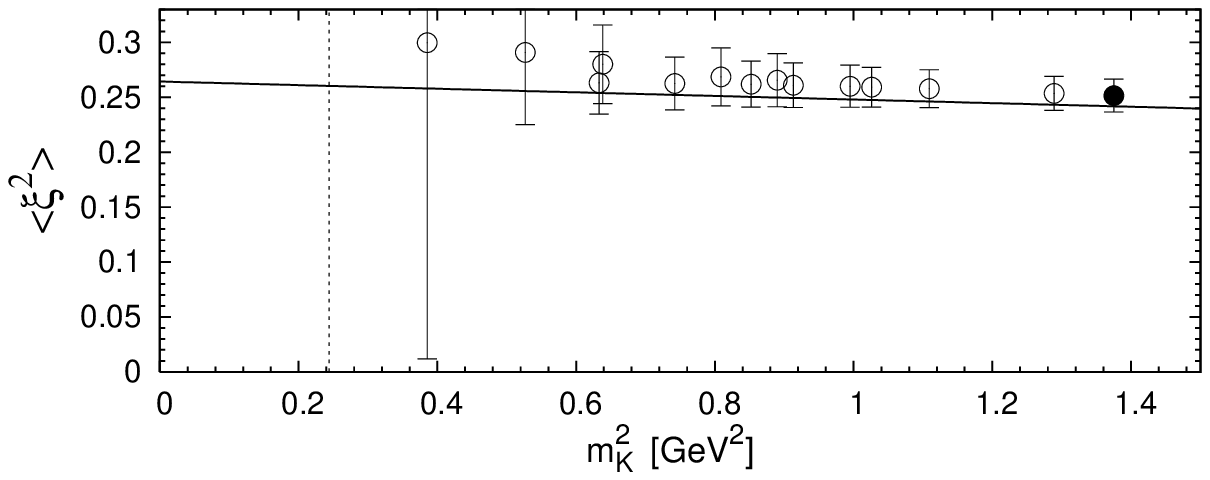}
    \includegraphics[width=0.8\hsize]{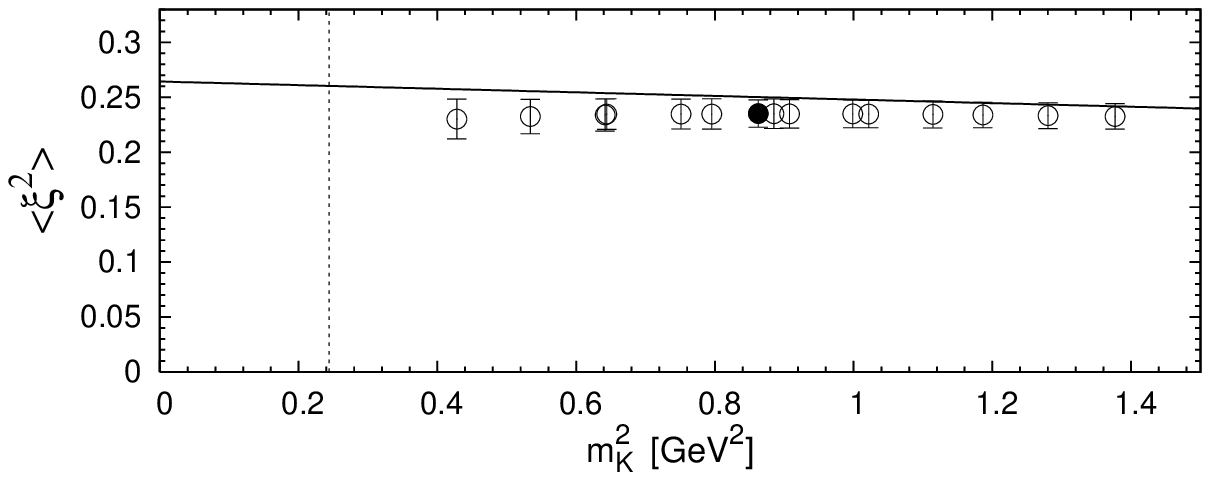}
    \includegraphics[width=0.8\hsize]{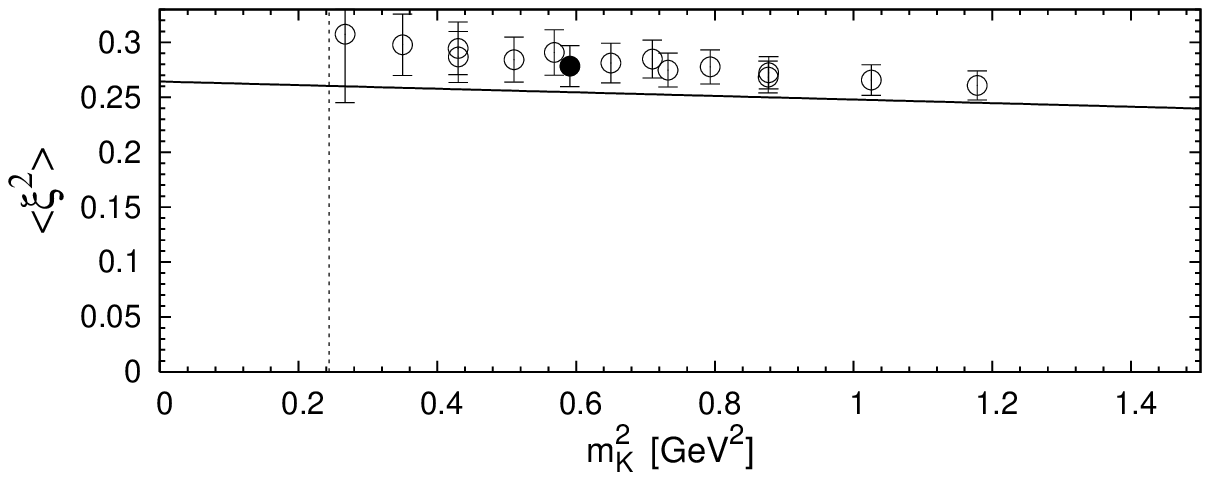}
    \includegraphics[width=0.8\hsize]{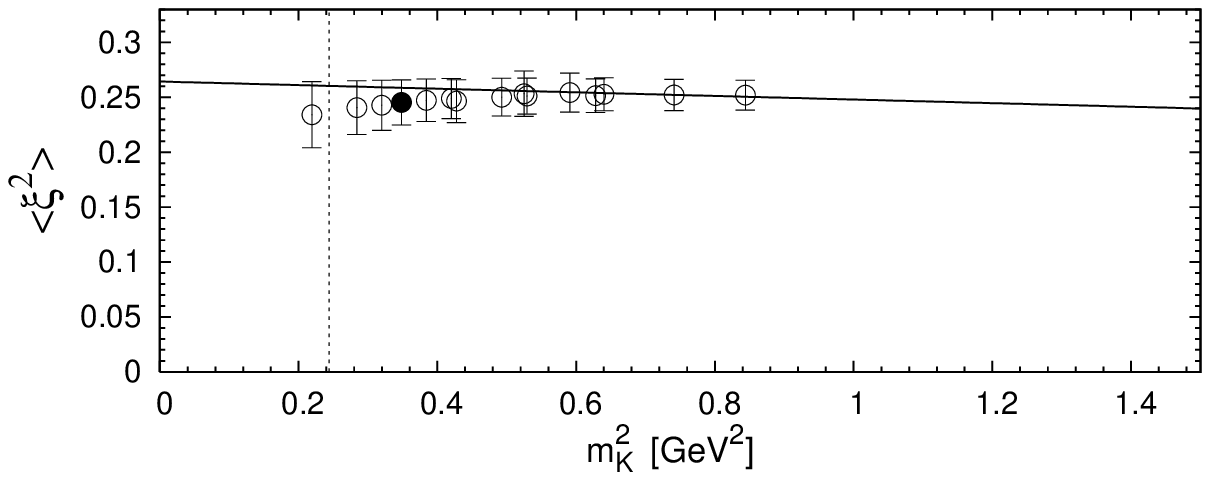}
    \caption{Results for $\langle\xi^2\rangle_K$ as in
      Fig.~\ref{fig:b5p29_2a5} but with all four values of
      $\kappa_\text{sea}$ separated for clarity.  Each solid line
      corresponds to the fitted ansatz, Eq.~(\ref{eq:xi2-fit}), for
      $m_\pi(\kappa_\text{sea},\,\kappa_\text{sea})$ evaluated at,
      going from top to bottom, $\kappa_\text{sea}=0.13400$,
      $\kappa_\text{sea}=0.13500$, $\kappa_\text{sea}=0.13550$ and
      $\kappa_\text{sea}=0.13590$. The vertical dotted lines
      correspond to the physical Kaon mass.}
    \label{fig:b5p29_global_a5}
  \end{center}
\end{figure}

In Fig.~\ref{fig:b5p29_2a5} we display all our results for
$\langle\xi^2\rangle_K$ for all four sea quark masses, together with the
fitted ansatz, Eq.~(\ref{eq:xi2-fit}), at the physical pion mass, i.e.
$\alpha + \beta m^2_{\pi,\text{phys}} + \gamma
m_K^2(\kappa_\text{val1},\,\kappa_\text{val2})$, given by the solid
line.
For further clarification, the result of this fit is also shown in
Fig.~\ref{fig:b5p29_global_a5} for each value of $\kappa_\text{sea}$
separately. 
In this figure, each solid line corresponds to the fitted ansatz,
Eq.~(\ref{eq:xi2-fit}), for
$m_\pi(\kappa_\text{sea},\,\kappa_\text{sea})$ evaluated at, going
from top to bottom, $\kappa_\text{sea}=0.13400$,
$\kappa_\text{sea}=0.13500$, $\kappa_\text{sea}=0.13550$ and
$\kappa_\text{sea}=0.13590$.  
For example, in the top figure, the solid line refers to $\alpha +
\beta m^2_{\pi}(0.13400,0.13400) + \gamma
m_K^2(\kappa_\text{val1},\,\kappa_\text{val2})$, where
$m^2_{\pi}(0.13400,0.13400)$ is taken from Table~\ref{tab:parameters}.

To obtain our final result, we insert the physical values for $m_\pi$
and $m_K$, together with the fitted parameters in
Eq.~(\ref{eq:xi2-fit-results}), into Eq.~(\ref{eq:xi2-fit}) and we
find in the \msbar  scheme at $\mu^2=4$~GeV$^2$
\be
\langle\xi^2\rangle_K^{\overline{\rm MS}}(\mu^2=4\,{\rm GeV}^2) =
0.260(6) \ .
\label{eq:xi2K-result}
\ee
Since we only have results with non-degenerate quark masses at one
value of $\beta=5.29$, we are not able to perform a continuum
extrapolation of $\langle\xi^2\rangle_K$.
We are, however, able to gain an estimate of the systematic error due
to discretisation effects by comparing the result for
$\langle\xi^2\rangle_\pi$ at $\beta=5.29$ with that in the continuum
limit (\ref{eq:xi2pi-result}).
Such a comparison suggests that there is a systematic error of roughly
$6\%$ due to discretisation effects.

Comparing the results in Eqs.~(\ref{eq:xi2pi-result}) and
(\ref{eq:xi2K-result}), we see that second moments for the Kaon
and pion coincide within errors, in agreement with findings in
Refs.~\cite{Ball:2006wn,Khodjamirian:2004ga}.

\subsubsection{First moment}

\begin{figure}[tb]
  \begin{center}
    \includegraphics[width=0.9\hsize]{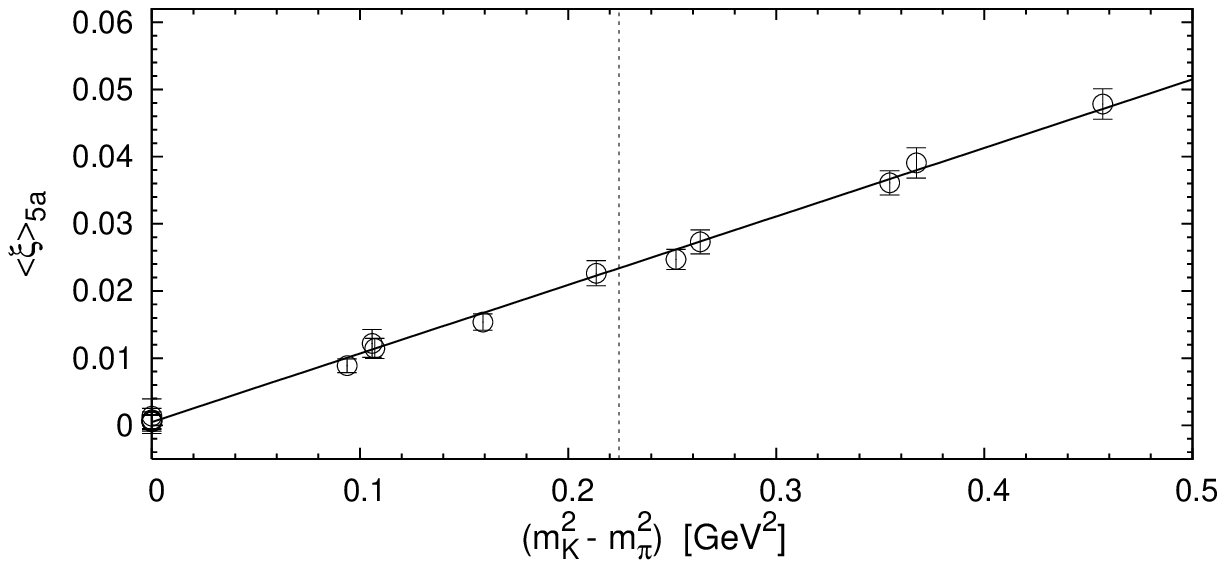}
    \includegraphics[width=0.9\hsize]{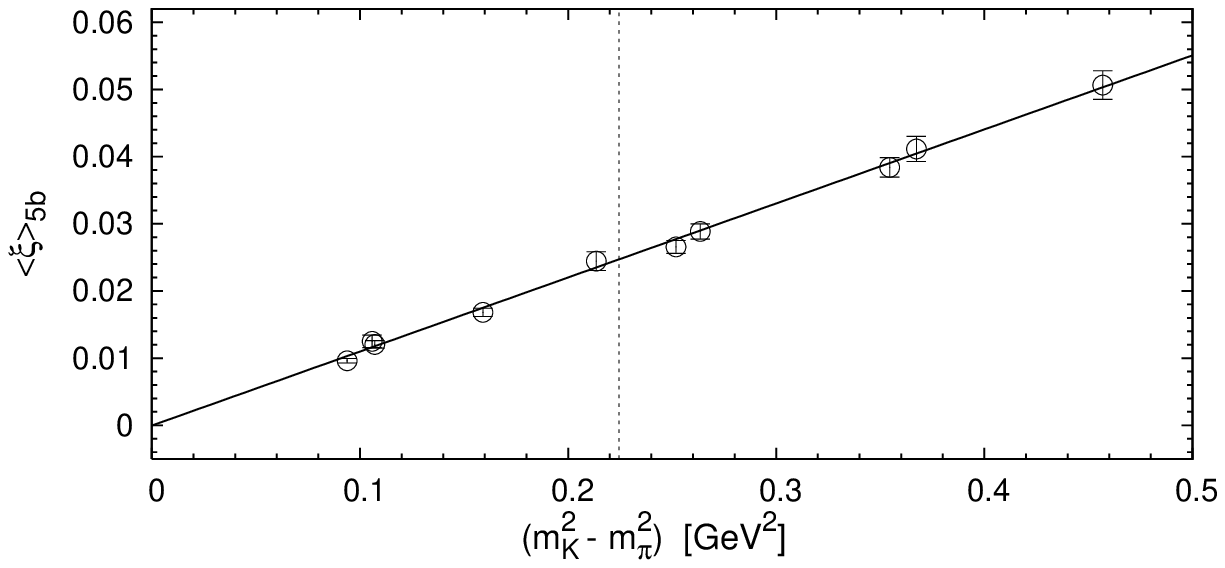}
    \caption{Result for $\langle\xi\rangle_a^5$ and
      $\langle\xi\rangle_b^5$ for $\beta=5.29,\ \kappa_{\rm
        sea}=0.13500$, in the \msbar scheme at
      $\mu^2=4\,\text{GeV}^2$. The vertical dotted line corresponds to
      the physical $m_K^2-m_\pi^2$ mass difference.}
    \label{fig:b5p29kp13500_1a5}
  \end{center}
\end{figure}

\begin{figure}[tb]
  \begin{center}
    \includegraphics[width=0.9\hsize]{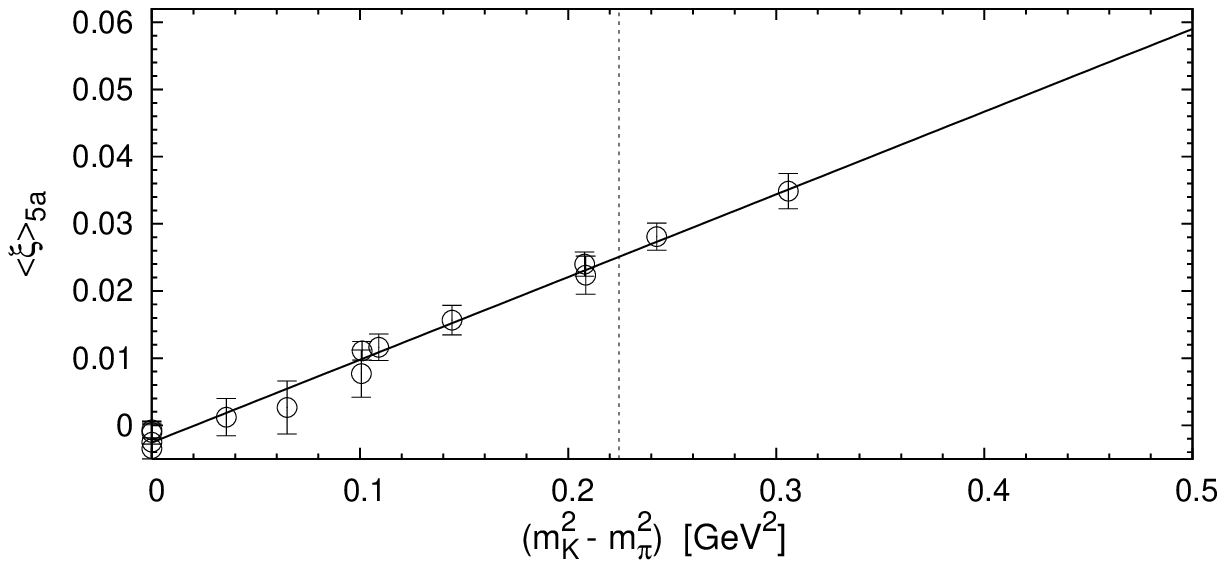}
    \includegraphics[width=0.9\hsize]{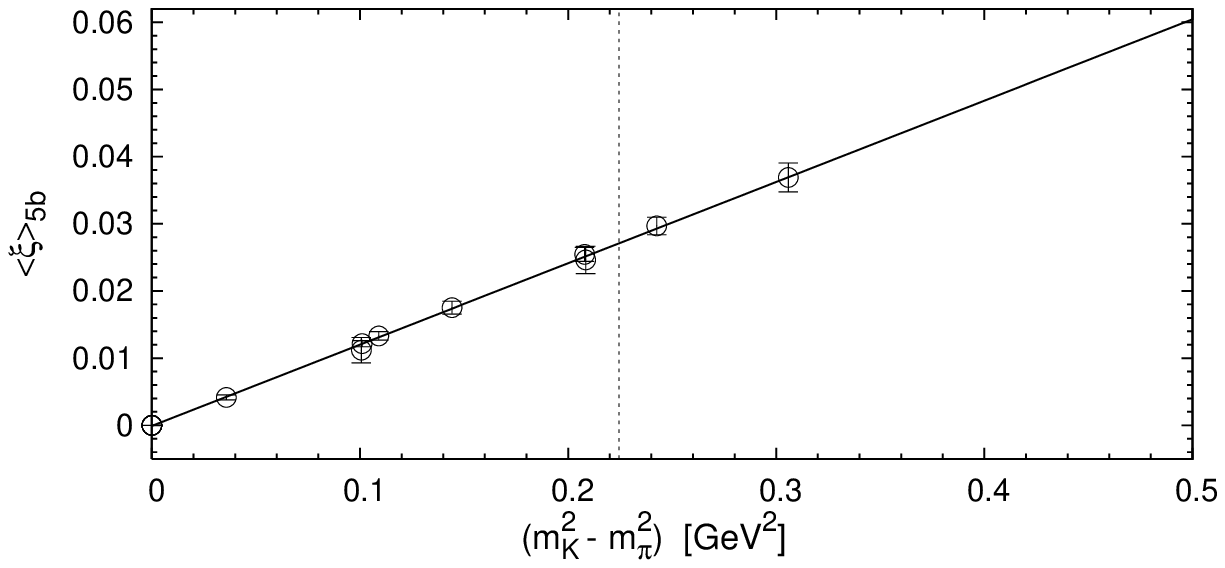}
    \caption{Result for $\langle\xi\rangle_a^5$ and 
      $\langle\xi\rangle_b^5$ for $\beta=5.29,\ \kappa_{\rm
        sea}=0.13590$, in the \msbar scheme at
      $\mu^2=4\,\text{GeV}^2$. The vertical dotted line corresponds to
      the physical $m_K^2-m_\pi^2$ mass difference.}
    \label{fig:b5p29kp13590_1a5}
  \end{center}
\end{figure}

\begin{table*}[tb]
  \begin{center}
    \caption{$\langle\xi\rangle_K$ at the physical $m_K^2 - m_\pi^2$
      mass splitting together with the slope, $B$, of the fit in
      Eq.~(\ref{eq:xi-fit}) for $\beta=5.29$, in the \msbar scheme at
      $\mu^2=4\,\text{GeV}^2$.
      The last row contains the values for $\langle\xi\rangle_K$
      chirally extrapolated in the sea quark mass to the physical
      point.    \label{table:xi-fit}}
   \begin{ruledtabular}
    \begin{tabular}{ccccccccc}
        $\kappa_{\rm sea}$ 
& $\langle\xi\rangle_a^{5}$ & $B_a^{5}$ 
& $\langle\xi\rangle_a^{45}$  & $B_a^{45}$
& $\langle\xi\rangle_b^{5}$ & $B_b^{5}$
& $\langle\xi\rangle_b^{45}$  & $B_b^{45}$    \\ \hline
    0.13400 & 0.0215(5) & 0.098(4) & 0.0222(8) & 0.099(6) & 0.0231(4) & 0.104(3) & 0.0223(19) & 0.121(18)   \\
    0.13500 & 0.0234(2) & 0.102(1) & 0.0231(2) & 0.102(1) & 0.0247(2) & 0.110(1) & 0.0248(1)  & 0.110(1)   \\
    0.13550 & 0.0240(5) & 0.120(3) & 0.0246(3) & 0.121(2) & 0.0276(3) & 0.123(2) & 0.0277(3)  & 0.124(2)   \\
    0.13590 & 0.0251(6) & 0.123(4) & 0.0237(4) & 0.126(3) & 0.0271(1) & 0.121(1) & 0.0267(2)  & 0.119(1)   \\
            & 0.0261(3) &          & 0.0252(11)&          & 0.0287(9) &          & 0.0289(16) &
     \end{tabular}
    \end{ruledtabular}
  \end{center}
\end{table*}

Figures~\ref{fig:b5p29kp13500_1a5} and \ref{fig:b5p29kp13590_1a5} show
the first moment, $\langle \xi\rangle_K$, for the working points
$\beta=5.29, \kappa_{\text{sea}}=0.13500$ and $\beta=5.29,
\kappa_{\text{sea}}=0.13590$, respectively, as obtained from the two
operators ${\cal O}^a_{\mu\nu}$ (\ref{eq:Oa41}) and ${\cal
  O}^b_{\mu\mu}$ (\ref{eq:Ob44}).
The results are plotted as a function of the mass splitting
of the two quarks making up the meson, or more specifically $m_K^2 -
m_\pi^2$.
Here $m_K$ is the mass of a pseudoscalar meson constructed
with one heavy and one light quark, while $m_\pi$ is the mass of a
pseudoscalar meson constructed with two light quarks, i.e.
$m_K(\kappa_\text{val1},\,\kappa_\text{val2}),\ 
m_\pi(\kappa_\text{val1},\,\kappa_\text{val1})$ with
$\kappa_\text{val1}\ge\kappa_\text{val2}$.
The points lie on a straight line, once again as predicted in
Ref.~\cite{Chen:2003fp,Chen:2005js}.

The vertical lines in Figs.~\ref{fig:b5p29kp13500_1a5} and
\ref{fig:b5p29kp13590_1a5} show the location of the physical
$K$-$\pi$ mass splitting and it is here that we extract our results for
$\langle\xi\rangle_K$ at each sea quark mass.
These results are given in Table~\ref{table:xi-fit} together with slopes
obtained from the simple fit
\be
\langle\xi\rangle_K = B(m_K^2 - m_\pi^2)\, .
\label{eq:xi-fit}
\ee
We observe that at each $\kappa_\text{sea}$, the four sets of results
obtained with two different operators and two different Kaon
interpolating fields all agree well.

\begin{figure}[tb]
  \begin{center}
    \includegraphics[width=0.9\hsize]{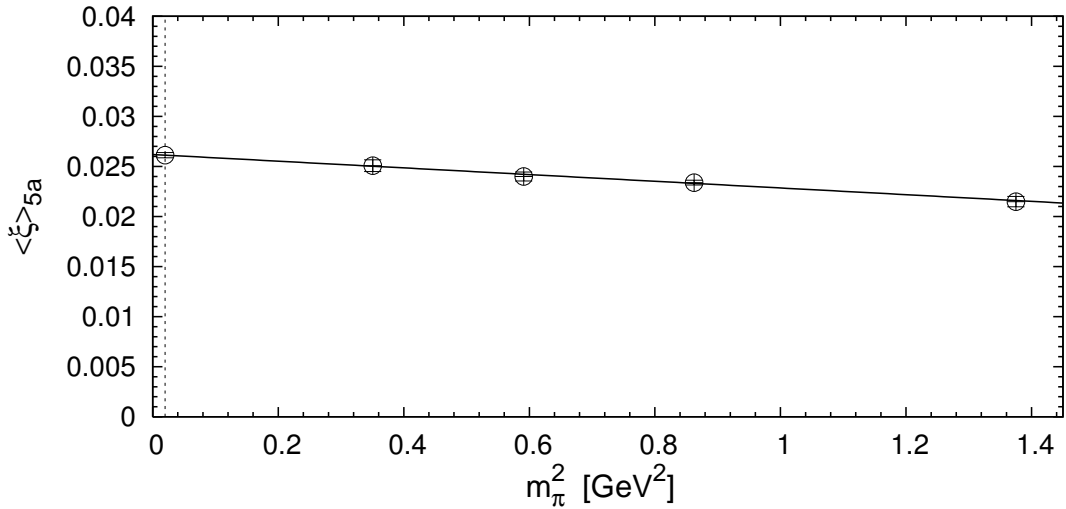}
    \caption{Sea quark mass dependence of $\langle\xi\rangle_a^5$ for
      $\beta=5.29$, in the \msbar scheme at $\mu^2=4\,\text{GeV}^2$.
      The vertical dotted lines correspond to the physical $\pi$ mass.}
    \label{fig:slope_1a5}
  \end{center}
\end{figure}

\begin{figure}[tb]
  \begin{center}
    \includegraphics[width=0.9\hsize]{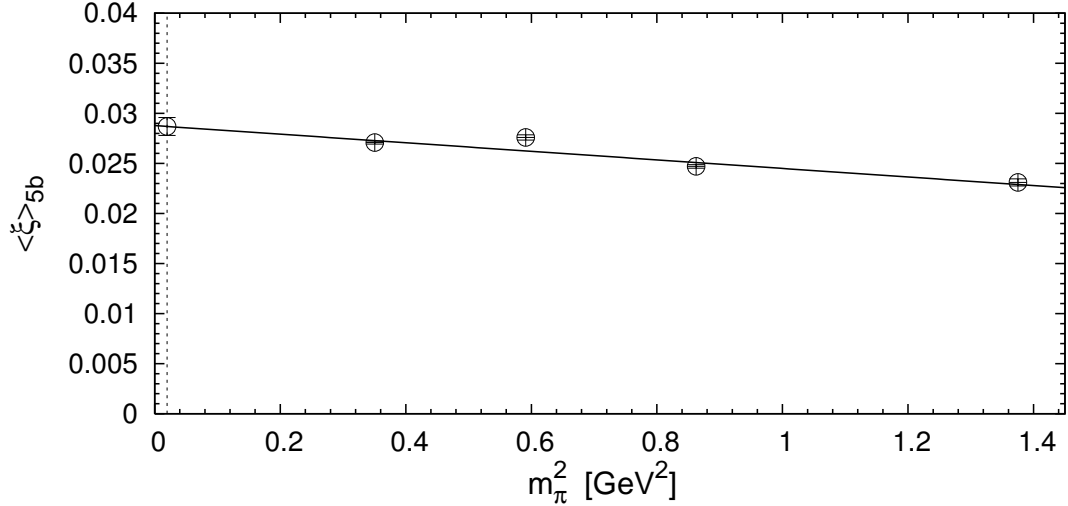}
    \caption{Sea quark mass dependence of $\langle\xi\rangle_b^5$ for
      $\beta=5.29$, in the \msbar scheme at
      $\mu^2=4\,\text{GeV}^2$. The vertical dotted lines
      correspond to the physical $\pi$ mass.}
    \label{fig:slope_1b5}
  \end{center}
\end{figure}

In order to extract a result at the physical pion mass, we examine the
sea quark mass dependence of our results by plotting them as a
function of the pion mass calculated with
$\kappa_\text{val}=\kappa_\text{sea}$ (Table~\ref{tab:parameters}) in
Figs.~\ref{fig:slope_1a5} and \ref{fig:slope_1b5} for the operators
${\cal O}^a_{\mu\nu}$ (\ref{eq:Oa41}) and ${\cal O}^b_{\mu\mu}$
(\ref{eq:Ob44}), respectively.
We extrapolate linearly in the mass of the light quark to the physical
pion mass and quote the results in the last row of
Table~\ref{table:xi-fit}.
Averaging over the four results, we find
\be
\langle\xi\rangle^{\overline{\rm MS}}_K(\mu^2=4\,{\rm GeV}^2) =
0.0272(5)\ .
\label{eq:xi-chiral}
\ee

Similar to the result for $\langle\xi^2\rangle_K$ in
Eq.~(\ref{eq:xi2K-result}), we expect that there is a systematic error
of roughly $6\%$ due to discretisation effects.

%
%

\section{Summary and Conclusions}
\label{sec:summary-conclusions}

We have presented results for the second moment of the pion's
distribution amplitide and the first two moments of the Kaon's
distribution amplitude, calculated on lattices generated by the
QCDSF/UKQCD collaboration with two flavours of dynamical fermions.  
We use nonperturbatively determined renormalization coefficients
(apart from the mixing with the operators containing total
derivatives, which is calculated perturbatively) to convert our result
to the $\overline{\rm MS}$ scheme at 4 GeV$^2$. 
Our results give model-independent insights into the distribution
amplitude of pseudoscalar mesons with degenerate and non-degenerate
quark masses.

We find for the pion $\langle \xi^2 \rangle_\pi =0.269(39)$, which is
in agreement with other results appearing in the literature and larger
than the asymptotic value. 
For the $K$-meson we obtain $\langle \xi \rangle_K =0.0272(5)(17)$ and
$\langle \xi^2 \rangle_K =0.260(6)(16)$, where the first error is
statistical and the second is an estimate of the systematic error due
to the fact that we have results with non-degenerate quarks at one
value of $\beta=5.29$ only, i.e., no continuum extrapolation.

\begin{figure}[tb]
  \begin{center}
    \includegraphics[width=0.9\hsize]{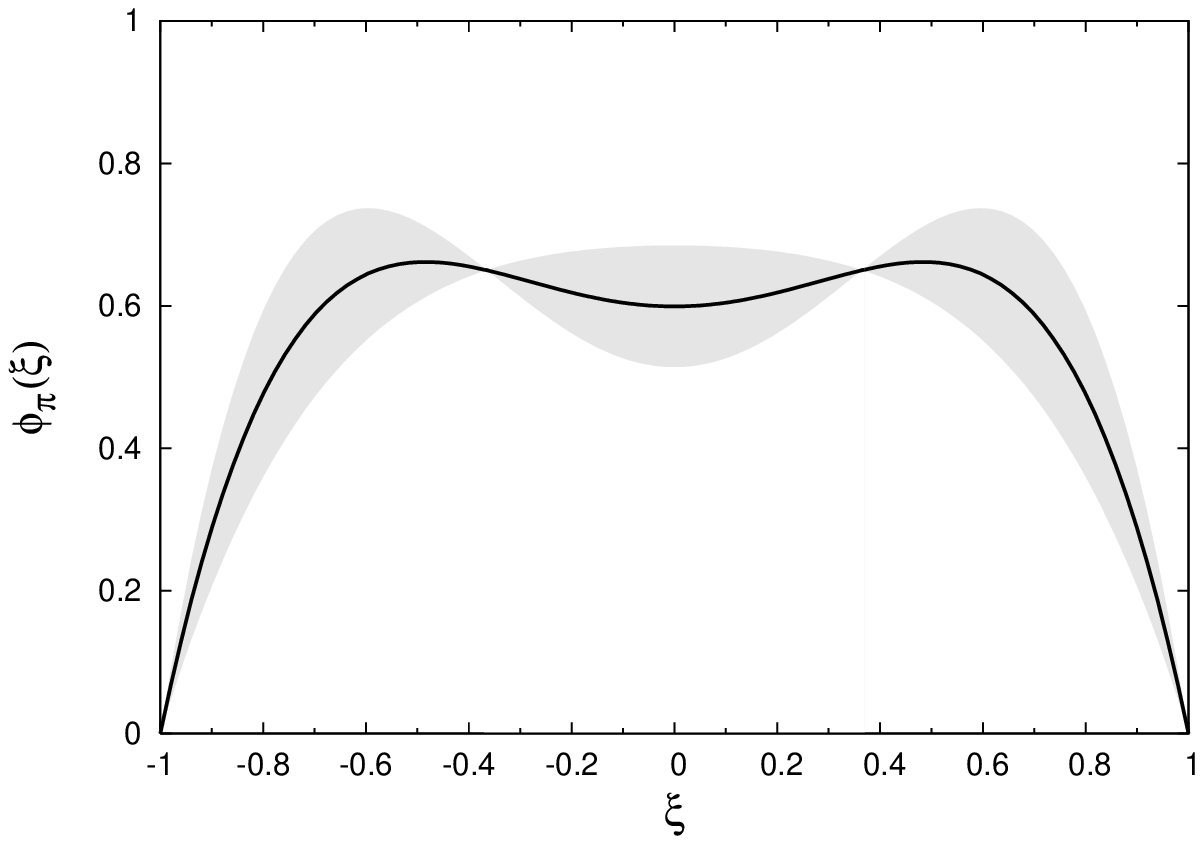}
    \caption{Distribution amplitude of the pion using the expansion in
      Eq.~(\ref{eq:gegen1}) with our result for $a_2^\pi=0.201(114)$ and
      $a_4^\pi=0$. This result is obtained in the $\overline{\rm MS}$
      scheme at $\mu^2=4\,\text{GeV}^2$.  The shaded area indicates
      the results obtained when $a_2^\pi$ varies between the maximum and
      minimum values allowed by its error.}
    \label{fig:phi_pi_a4=0}
  \end{center}
\end{figure}

\begin{figure}[tb]
  \begin{center}
    \includegraphics[width=0.9\hsize]{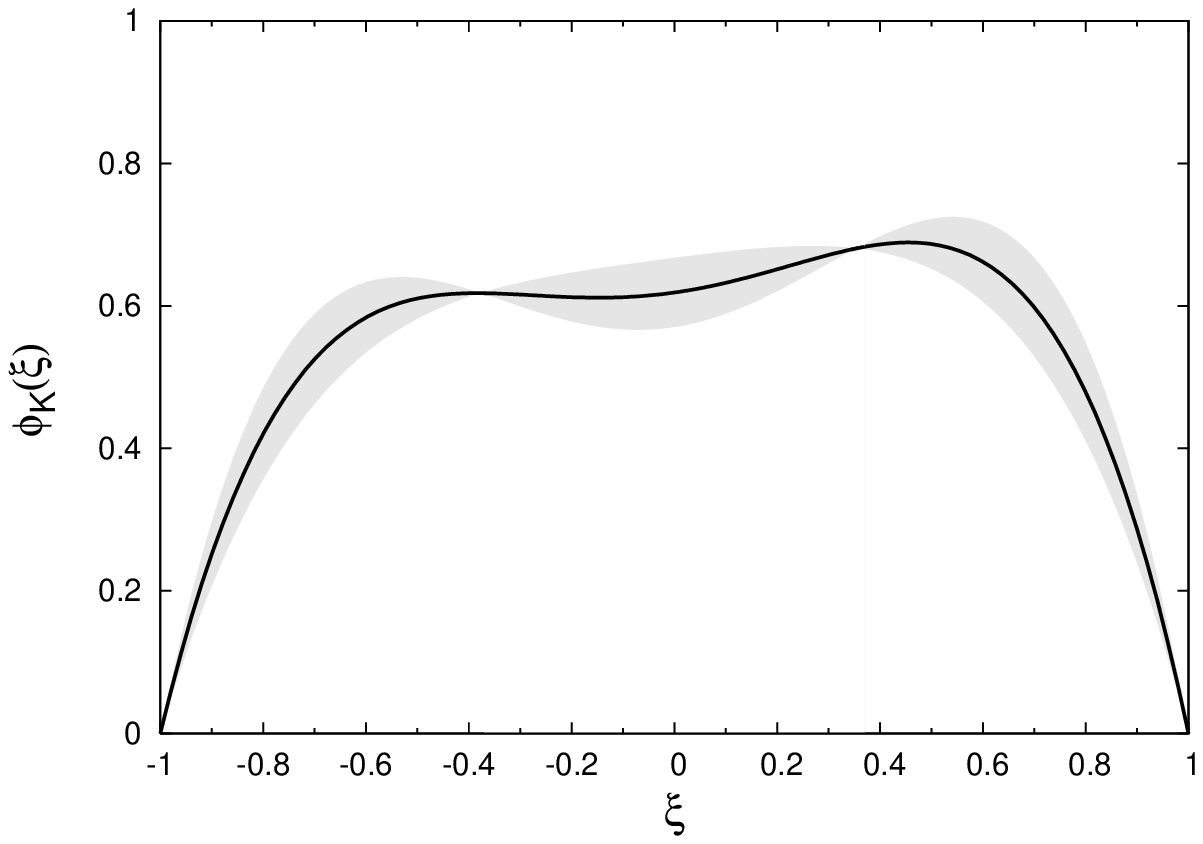}
    \caption{Distribution amplitude of the Kaon using the expansion in
      Eq.~(\ref{eq:gegen1}) with our results for $a_1^K=0.0453(9)(29)$ and
      $a_2^K=0.175(18)(47)$.  These results are obtained in the $\overline{\rm
        MS}$ scheme at $\mu^2=4\,\text{GeV}^2$.  The shaded area
      indicates the results obtained when $a_1^K$ and $a_2^K$ vary between
      the maximum and minimum values allowed by their errors.}
    \label{fig:phi_K}
  \end{center}
\end{figure}

The coefficients $a_n$ in the Gegenbauer expansion of the DAs in
Eq.~(\ref{eq:gegen1}) are related to the moments $\langle \xi^n\rangle$
by simple agebraic relations (\ref{eq:geg-alg}).
Using our result in Eq.~(\ref{eq:xi2pi-result}) we obtain, for the
$\pi$-meson
\begin{equation}
   a_2^\pi (\mu^2=4~\mbox{GeV}^2) = 0.201(114)\,, 
\end{equation}
and from Eqs.~(\ref{eq:xi2K-result}) and (\ref{eq:xi-chiral}) for
the $K$-meson:
\begin{eqnarray}
   a_1^K (\mu^2&=&4~\mbox{GeV}^2) = 0.0453(9)(29)\,, 
\\
   a_2^K (\mu^2&=&4~\mbox{GeV}^2) = 0.175(18)(47)\,.
\end{eqnarray}
While our result for $a_2^\pi$ is larger than the transverse lattice
result \cite{Dalley:2002nj}, all three numbers are well within the
range suggested by QCD sum rule estimates and supported (for the pion)
by the analysis of CLEO data on the $\pi\gamma^*\gamma$ transition
form factor, cf. Eqs.~(\ref{a2pi}) and (\ref{a1K}).
Also the $SU(3)$ breaking in the second Gegenbauer coefficient turns
out to be small, in agreement with
\cite{Ball:2006wn,Khodjamirian:2004ga}.
We note that in the context of SU(3) flavour violation, one might be
worried about the absence of a dynamical strange quark in our
simulations, however there has recently appeared a $N_f=2+1$ lattice
calculation of $a_1^K$ \cite{Boyle:2006pw} which is in good
agreement with our result, giving us confidence that the effects of a
dynamical strange quark are probably small.

Our results indicate that it is important to consider not only the
chiral extrapolation of the lattice results to the physical quark
masses, but also to perform simulations at small enough lattice
spacings to allow for a reliable extrapolation to the continuum limit.

The corresponding DAs obtained by the truncation of the general
expression in Eq.~(\ref{eq:gegen1}) after the second term are shown in
Fig.~\ref{fig:phi_pi_a4=0} and Fig.~\ref{fig:phi_K} for the $\pi$ and
the $K$-mesons, respectively.
Note that the $K$-meson DA is tilted towards larger momentum fractions
carried by the heavier strange quark, which is in agreement with
general expectations.

\begin{figure}[tb]
  \begin{center}
    \includegraphics[width=0.9\hsize]{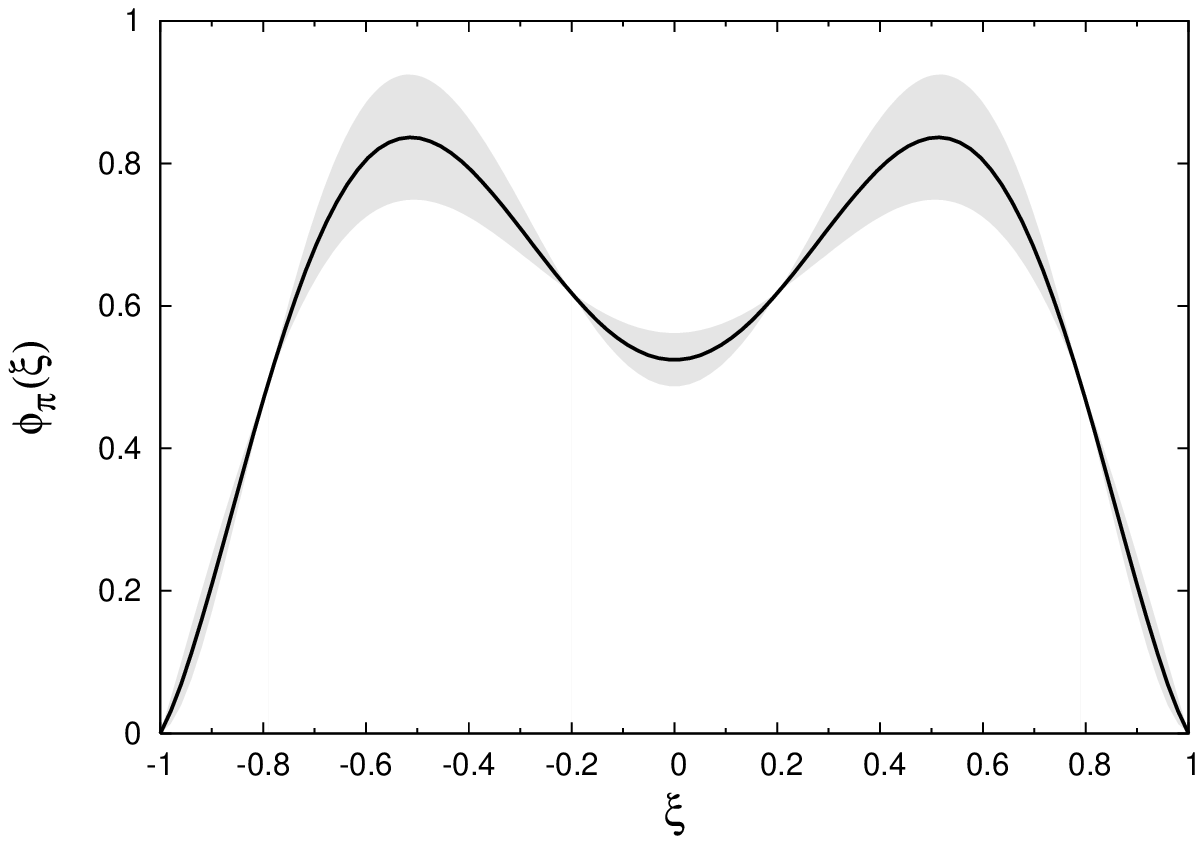}
    \caption{Distribution amplitude of the pion using the expansion in
      Eq.~(\ref{eq:gegen1}) with $a_2^\pi=0.201(114)$ and
      $a_4^\pi=-0.10(5)$.  This result is obtained in the
      $\overline{\rm MS}$ scheme at $\mu^2=4\,\text{GeV}^2$. The
      shaded area indicates the results obtained when we fix $a_2^\pi$
      at its central value and let $a_4^\pi$ vary between the maximum
      and minimum values allowed by its error.}
    \label{fig:phi_pi_a4=0.1}
  \end{center}
\end{figure}

In order to illustrate the possible effect of higher-order terms in
the Gegenbauer expansion, we also show in Fig.~\ref{fig:phi_pi_a4=0.1}
the pion DA obtained with the addition of the fourth-order polynomial
with the coefficient $a_4^\pi = -0.10(5)$ taken from
Ref.~\cite{Bakulev:2005cp}.
In both cases (with and without $a_4^\pi$) the value of the DA in the
middle point agrees well with the estimate in Eq.~(\ref{eq:phi-mid}).  
The question whether the ``camel-hump'' structure of the DA is present
in the physical DA depends on the contribution of yet higher-order
polynomials that are beyond the reach of the present analysis.

%
%

\begin{acknowledgments}
  We thank Alan Irving for providing $r_0/a$.  The
  numerical calculations have been done on the Hitachi SR8000 at LRZ
  (Munich), on the Cray T3E at EPCC (Edinburgh) under PPARC grant
  PPA/G/S/1998/00777 \cite{Allton:2001sk} and on the APE{\it 1000} and
  apeNEXT at NIC/DESY (Zeuthen). We thank the operating staff for
  support.  This work was supported in part by the DFG (Forschergruppe
  Gitter-Hadronen-Ph{\"a}nomenologie) and in part by the EU Integrated
  Infrastructure Initiative Hadron Physics (I3HP) under contract
  number RII3-CT-2004-506078. W.S.~acknowledges funding by the
  Alexander-von-Humboldt foundation and thanks the Physics Department
  of the National Taiwan University for their hospitality. J.Z. would
  like to thank Andreas J\"uttner for useful discussions. W.S.~also
  thanks Jiunn-Wei Chen for valuable remarks and discussions.
\end{acknowledgments}

%
%

\appendix

\section{\label{sec:latt-results-work}Lattice results by working
  point}
The following tables summarize our findings at individual working
points. 

\begin{table*}
\caption{Bare results for $\langle\xi\rangle$ and $\langle\xi^2\rangle$ for various
  choices of valences quarks with
  $\kappa_\text{val1}\ge\kappa_\text{val2}$, together with the
  corresponding pseudoscalar mass,
  $m_\text{ps}(\kappa_\text{val1},\,\kappa_\text{val2})$. Here
  $\beta=5.29$ and $\kappa_\text{sea}=0.13400$.}
\label{table:b5p29kp13400}
\begin{ruledtabular}
\begin{tabular}{cc|ccccccc}
\multicolumn{9}{c}{$\beta = 5.29,\ \kappa_{\rm sea}=0.13400$ } \\ \hline
$\kappa_{\rm val1}$ & $\kappa_{\rm val2}$ 
& $am_{\rm ps}$
& $\langle\xi\rangle_a^5$
& $\langle\xi\rangle_a^{45}$
& $\langle\xi\rangle_b^5$
& $\langle\xi\rangle_b^{45}$
& $\langle\xi^2\rangle_a^{5}$
& $\langle\xi^2\rangle_a^{45}$
\\ \hline  \hline 
0.13400 & 0.13400 & 0.5767(11) & -0.00027(99) & 0.0001(11) &              &              & 0.1434(54) & 0.1537(47) \\
0.13440 & 0.13400 & 0.5583(10) &  0.0053(10)  & 0.0059(11) & 0.005873(71) & 0.005989(92) & 0.1445(58) & 0.1552(50) \\
0.13525 & 0.13400 & 0.5179(11) &  0.0179(14)  & 0.0190(15) & 0.01844(30)  & 0.01888(40)  & 0.1468(70) & 0.1586(60) \\ 
0.13525 & 0.13440 & 0.4981(11) &  0.0121(14)  & 0.0131(15) & 0.01257(24)  & 0.01290(32)  & 0.1475(76) & 0.1600(64) \\
0.13525 & 0.13525 & 0.4541(11) &              &            &              &              & 0.1490(94) & 0.1636(77) \\
0.13580 & 0.13400 & 0.4906(11) &  0.0259(15)  & 0.0270(17) & 0.02666(58)  & 0.02741(78)  & 0.1480(83) & 0.1607(70) \\
0.13580 & 0.13440 & 0.4700(11) &  0.0201(16)  & 0.0210(18) & 0.02080(53)  & 0.02146(73)  & 0.1485(90) & 0.1621(75) \\
0.13580 & 0.13525 & 0.4237(12) &  0.0076(19)  & 0.0083(20) & 0.00826(33)  & 0.00867(50)  & 0.149(11)  & 0.1656(90) \\
0.13580 & 0.13580 & 0.3913(14) &              &            &              &              & 0.150(14)  & 0.169(11)  \\
0.13630 & 0.13400 & 0.4639(15) &  0.0332(19)  & 0.0337(21) & 0.0352(14)   & 0.0367(19)   & 0.151(11)  & 0.1633(91) \\
0.13630 & 0.13440 & 0.4423(16) &  0.0275(19)  & 0.0277(22) & 0.0294(15)   & 0.0311(20)   & 0.153(13)  & 0.1651(99) \\ 
0.13630 & 0.13525 & 0.3929(21) &  0.0154(23)  & 0.0150(25) & 0.0172(16)   & 0.0191(27)   & 0.159(18)  & 0.171(13)  \\
0.13630 & 0.13580 & 0.3569(32) &  0.0081(28)  & 0.0072(29) & 0.0093(18)   & -0.0009(17)  & 0.165(35)  & 0.178(18)  \\
0.13630 & 0.13630 & 0.305(14)  &              &            &              &              & 0.17(15)   & 0.183(36)  \\
\end{tabular}
\end{ruledtabular}
\end{table*}

\begin{table*}
\caption{Bare results for $\langle\xi\rangle$ and $\langle\xi^2\rangle$ for various
  choices of valences quarks with
  $\kappa_\text{val1}\ge\kappa_\text{val2}$, together with the
  corresponding pseudoscalar mass,
  $m_\text{ps}(\kappa_\text{val1},\,\kappa_\text{val2})$. Here
  $\beta=5.29$ and $\kappa_\text{sea}=0.13500$.}
\label{table:b5p29kp13500}
\begin{ruledtabular}
\begin{tabular}{cc|ccccccc}
\multicolumn{9}{c}{$\beta = 5.29,\ \kappa_{\rm sea}=0.13500$ } \\ \hline
$\kappa_{\rm val1}$ & $\kappa_{\rm val2}$ 
& $am_{\rm ps}$
& $\langle\xi\rangle_a^5$
& $\langle\xi\rangle_a^{45}$
& $\langle\xi\rangle_b^5$
& $\langle\xi\rangle_b^{45}$
& $\langle\xi^2\rangle_a^{5}$
& $\langle\xi^2\rangle_a^{45}$
\\ \hline  \hline 
0.13390 & 0.13390 & 0.53134(84) & 0.00035(68) & 0.00028(78) &              &              & 0.1332(29) & 0.1521(32) \\
0.13430 & 0.13390 & 0.51236(86) & 0.00622(70) & 0.00647(81) & 0.006642(76) & 0.006706(75) & 0.1336(31) & 0.1532(34) \\
0.13430 & 0.13430 & 0.49293(76) & 0.00040(73) & 0.00028(85) &              &              & 0.1339(28) & 0.1544(28) \\ 
0.13500 & 0.13390 & 0.47803(88) & 0.01731(94) & 0.01713(92) & 0.01838(28)  & 0.01850(27)  & 0.1342(35) & 0.1555(38) \\
0.13500 & 0.13430 & 0.45771(90) & 0.01078(80) & 0.01100(96) & 0.01163(20)  & 0.01179(20)  & 0.1344(37) & 0.1566(40) \\
0.13500 & 0.13500 & 0.42053(82) & 0.00049(92) & 0.0002(10)  &              &              & 0.1346(37) & 0.1587(35) \\
0.13550 & 0.13390 & 0.45254(91) & 0.0253(11)  & 0.02527(98) & 0.02657(50)  & 0.02686(51)  & 0.1345(40) & 0.1573(43) \\
0.13550 & 0.13430 & 0.43142(93) & 0.0191(11)  & 0.0188(10)  & 0.01996(45)  & 0.02013(45)  & 0.1345(42) & 0.1587(45) \\
0.13550 & 0.13500 & 0.39249(98) & 0.0080(10)  & 0.0078(12)  & 0.00833(27)  & 0.00832(27)  & 0.1344(47) & 0.1602(50) \\
0.13550 & 0.13550 & 0.36283(91) & 0.0006(12)  & 0.0001(13)  &              &              & 0.1344(50) & 0.1619(44) \\
0.13600 & 0.13390 & 0.42603(97) & 0.0335(13)  & 0.0335(15)  & 0.03505(92)  & 0.03518(93)  & 0.1346(46) & 0.1590(49) \\ 
0.13600 & 0.13430 & 0.40392(99) & 0.0274(14)  & 0.0271(15)  & 0.02846(90)  & 0.02842(92)  & 0.1344(49) & 0.1598(52) \\
0.13600 & 0.13500 & 0.3628(10)  & 0.0159(12)  & 0.0157(13)  & 0.01690(79)  & 0.01658(86)  & 0.1339(56) & 0.1617(58) \\
0.13600 & 0.13550 & 0.3309(11)  & 0.0085(14)  & 0.0079(15)  & 0.00864(57)  & 0.00827(71)  & 0.1332(64) & 0.1637(64) \\
0.13600 & 0.13600 & 0.2962(11)  & 0.0010(18)  & 0.0006(19)  &              &              & 0.1320(81) & 0.1665(65) \\
\end{tabular}
\end{ruledtabular}
\end{table*}

\begin{table*}
\caption{Bare results for $\langle\xi\rangle$ and $\langle\xi^2\rangle$ for various
  choices of valences quarks with
  $\kappa_\text{val1}\ge\kappa_\text{val2}$, together with the
  corresponding pseudoscalar mass,
  $m_\text{ps}(\kappa_\text{val1},\,\kappa_\text{val2})$. Here
  $\beta=5.29$ and $\kappa_\text{sea}=0.13550$.}
\label{table:b5p29kp13550}
\begin{ruledtabular}
\begin{tabular}{cc|ccccccc}
\multicolumn{9}{c}{$\beta = 5.29,\ \kappa_{\rm sea}=0.13550$ } \\ \hline
$\kappa_{\rm val1}$ & $\kappa_{\rm val2}$ 
& $am_{\rm ps}$
& $\langle\xi\rangle_a^5$
& $\langle\xi\rangle_a^{45}$
& $\langle\xi\rangle_b^5$
& $\langle\xi\rangle_b^{45}$
& $\langle\xi^2\rangle_a^{5}$
& $\langle\xi^2\rangle_a^{45}$
\\ \hline  \hline 
0.13430 & 0.13430 & 0.46159(55) & -0.0006(8)  & -0.0005(10) &            &            & 0.1484(37)  & 0.1579(37) \\
0.13490 & 0.13430 & 0.43065(56) &  0.0109(9)  &  0.0110(12) & 0.0115(1)  & 0.0116(2)  & 0.1510(41)  & 0.1617(40) \\
0.13490 & 0.13490 & 0.39820(58) & -0.0008(11) & -0.0008(14) &            &            & 0.1525(45)  & 0.1645(45) \\ 
0.13550 & 0.13430 & 0.39837(59) &  0.0226(11) &  0.0228(14) & 0.0234(4)  & 0.0236(4)  & 0.1546(47)  & 0.1670(46) \\
0.13550 & 0.13490 & 0.36397(60) &  0.0106(14) &  0.0106(17) & 0.0118(2)  & 0.0120(3)  & 0.1559(52)  & 0.1698(52) \\
0.13550 & 0.13550 & 0.32696(64) & -0.0020(16) & -0.0019(20) &            &            & 0.1578(76)  & 0.1737(68) \\
0.13585 & 0.13430 & 0.37873(62) &  0.0293(13) &  0.0296(16) & 0.0304(6)  & 0.0306(8)  & 0.1574(52)  & 0.1713(53) \\
0.13585 & 0.13490 & 0.34288(64) &  0.0166(18) &  0.0166(21) & 0.0190(5)  & 0.0190(7)  & 0.1594(71)  & 0.1739(66) \\
0.13585 & 0.13550 & 0.30381(67) &  0.0045(19) &  0.0049(23) & 0.0071(3)  & 0.0071(6)  & 0.1610(88)  & 0.1791(78) \\
0.13585 & 0.13585 & 0.27895(72) & -0.0034(22) & -0.0026(27) &            &            & 0.162(10)   & 0.1841(91) \\
0.13620 & 0.13430 & 0.35835(74) &  0.0355(17) &  0.0362(21) & 0.0380(13) & 0.0382(17) & 0.1613(64)  & 0.1770(66) \\ 
0.13620 & 0.13490 & 0.32065(78) &  0.0220(24) &  0.0226(27) & 0.0266(13) & 0.0269(21) & 0.1646(88)  & 0.1812(83) \\
0.13620 & 0.13550 & 0.27880(88) &  0.0098(26) &  0.0112(30) & 0.0147(12) &            & 0.167(11)   & 0.1870(97) \\
0.13620 & 0.13585 & 0.2515(11)  &  0.0038(27) &  0.0051(33) &            &            & 0.168(13)   & 0.193(11)  \\
0.13620 & 0.13620 & 0.2198(26)  & -0.0081(56) & -0.0044(51) &            &            & 0.173(33)   & 0.201(18)  \\
\end{tabular}
\end{ruledtabular}
\end{table*}

\begin{table*}
\caption{Bare results for $\langle\xi\rangle$ and $\langle\xi^2\rangle$ for various
  choices of valences quarks with
  $\kappa_\text{val1}\ge\kappa_\text{val2}$, together with the
  corresponding pseudoscalar mass,
  $m_\text{ps}(\kappa_\text{val1},\,\kappa_\text{val2})$. Here
  $\beta=5.29$ and $\kappa_\text{sea}=0.13590$.}
\label{table:b5p29kp13590}
\begin{ruledtabular}
\begin{tabular}{cc|ccccccc}
\multicolumn{9}{c}{$\beta = 5.29,\ \kappa_{\rm sea}=0.13590$ } \\ \hline
$\kappa_{\rm val1}$ & $\kappa_{\rm val2}$ 
& $am_{\rm ps}$
& $\langle\xi\rangle_a^5$
& $\langle\xi\rangle_a^{45}$
& $\langle\xi\rangle_b^5$
& $\langle\xi\rangle_b^{45}$
& $\langle\xi^2\rangle_a^{5}$
& $\langle\xi^2\rangle_a^{45}$
\\ \hline  \hline 
0.13490 & 0.13490 & 0.37239(65) & -0.0005(8)  & -0.0021(10) &            &            & 0.1437(43) & 0.1605(38) \\
0.13530 & 0.13490 & 0.34894(67) &  0.0078(0)  &  0.0062(11) & 0.0084(1)  & 0.0084(2)  & 0.1438(48) & 0.1628(42) \\
0.13530 & 0.13530 & 0.32430(69) & -0.0008(12) & -0.0029(14) &            &            & 0.1441(54) & 0.1661(46) \\ 
0.13575 & 0.13490 & 0.32121(71) &  0.0168(12) &  0.0156(13) & 0.0176(4)  & 0.0176(6)  & 0.1434(57) & 0.1660(49) \\
0.13575 & 0.13530 & 0.29483(73) &  0.0081(14) &  0.0066(15) & 0.0092(3)  & 0.0092(4)  & 0.1432(65) & 0.1693(55) \\
0.13575 & 0.13575 & 0.26270(77) & -0.0018(20) & -0.0038(21) &            &            & 0.1420(79) & 0.1740(65) \\
0.13590 & 0.13490 & 0.31158(74) &  0.0197(13) &  0.0186(15) & 0.0205(6)  & 0.0204(8)  & 0.1449(74) & 0.1672(53) \\
0.13590 & 0.13530 & 0.28450(75) &  0.0110(15) &  0.0097(17) & 0.0121(5)  & 0.0120(7)  & 0.1427(71) & 0.1705(59) \\
0.13590 & 0.13575 & 0.25123(80) &  0.0008(19) & -0.0005(21) & 0.0029(2)  & 0.0028(3)  & 0.1411(86) & 0.1751(71) \\
0.13590 & 0.13590 & 0.23925(82) & -0.0024(25) & -0.0042(26) &            &            & 0.1401(94) & 0.1769(77) \\
0.13617 & 0.13490 & 0.29364(80) &  0.0244(17) &  0.0238(20) & 0.0255(12) & 0.0251(16) & 0.1447(93) & 0.1692(65) \\ 
0.13617 & 0.13530 & 0.26503(81) &  0.0157(19) &  0.0149(22) & 0.0170(13) & 0.0162(17) & 0.1407(87) & 0.1725(73) \\
0.13617 & 0.13575 & 0.22921(86) &  0.0054(25) &  0.0047(28) & 0.0077(13) &            & 0.139(11)  & 0.1770(88) \\
0.13617 & 0.13590 & 0.21603(89) &  0.0019(28) &  0.0013(30) &            &            & 0.138(12)  & 0.1787(96) \\
0.13617 & 0.13617 & 0.1897(10)  & -0.0042(44) & -0.0044(44) &            &            & 0.134(15)  & 0.182(12)  \\
\end{tabular}
\end{ruledtabular}
\end{table*}

%
%

\bibliography{PSDA_References}

\begin{thebibliography}{71}
\expandafter\ifx\csname natexlab\endcsname\relax\def\natexlab#1{#1}\fi
\expandafter\ifx\csname bibnamefont\endcsname\relax
  \def\bibnamefont#1{#1}\fi
\expandafter\ifx\csname bibfnamefont\endcsname\relax
  \def\bibfnamefont#1{#1}\fi
\expandafter\ifx\csname citenamefont\endcsname\relax
  \def\citenamefont#1{#1}\fi
\expandafter\ifx\csname url\endcsname\relax
  \def\url#1{\texttt{#1}}\fi
\expandafter\ifx\csname urlprefix\endcsname\relax\def\urlprefix{URL }\fi
\providecommand{\bibinfo}[2]{#2}
\providecommand{\eprint}[2][]{\url{#2}}

\bibitem[{\citenamefont{Brodsky and Lepage}(1989)}]{Brodsky:1989pv}
\bibinfo{author}{\bibfnamefont{S.~J.} \bibnamefont{Brodsky}} \bibnamefont{and}
  \bibinfo{author}{\bibfnamefont{G.~P.} \bibnamefont{Lepage}},
  \bibinfo{journal}{Adv. Ser. Direct. High Energy Phys.}
  \textbf{\bibinfo{volume}{5}}, \bibinfo{pages}{93}
 (\bibinfo{year}{1989}).

\bibitem[{\citenamefont{Chernyak and Zhitnitsky}(1977)}]{Chernyak:1977as}
\bibinfo{author}{\bibfnamefont{V.~L.} \bibnamefont{Chernyak}} \bibnamefont{and}
  \bibinfo{author}{\bibfnamefont{A.~R.} \bibnamefont{Zhitnitsky}},
  \bibinfo{journal}{JETP Lett.} \textbf{\bibinfo{volume}{25}},
  \bibinfo{pages}{510}
 (\bibinfo{year}{1977}).

\bibitem[{\citenamefont{Chernyak and Zhitnitsky}(1980)}]{Chernyak:1980dj}
\bibinfo{author}{\bibfnamefont{V.~L.} \bibnamefont{Chernyak}} \bibnamefont{and}
  \bibinfo{author}{\bibfnamefont{A.~R.} \bibnamefont{Zhitnitsky}},
  \bibinfo{journal}{Sov. J. Nucl. Phys.} \textbf{\bibinfo{volume}{31}},
  \bibinfo{pages}{544}
 (\bibinfo{year}{1980}).

\bibitem[{\citenamefont{Efremov and
  Radyushkin}(1980{\natexlab{a}})}]{Efremov:1979qk}
\bibinfo{author}{\bibfnamefont{A.~V.} \bibnamefont{Efremov}} \bibnamefont{and}
  \bibinfo{author}{\bibfnamefont{A.~V.} \bibnamefont{Radyushkin}},
  \bibinfo{journal}{Phys. Lett.} \textbf{\bibinfo{volume}{B94}},
  \bibinfo{pages}{245}
 (\bibinfo{year}{1980}{\natexlab{a}}).

\bibitem[{\citenamefont{Efremov and
  Radyushkin}(1980{\natexlab{b}})}]{Efremov:1978rn}
\bibinfo{author}{\bibfnamefont{A.~V.} \bibnamefont{Efremov}} \bibnamefont{and}
  \bibinfo{author}{\bibfnamefont{A.~V.} \bibnamefont{Radyushkin}},
  \bibinfo{journal}{Theor. Math. Phys.} \textbf{\bibinfo{volume}{42}},
  \bibinfo{pages}{97}
 (\bibinfo{year}{1980}{\natexlab{b}}).

\bibitem[{\citenamefont{Lepage and Brodsky}(1979)}]{Lepage:1979zb}
\bibinfo{author}{\bibfnamefont{G.~P.} \bibnamefont{Lepage}} \bibnamefont{and}
  \bibinfo{author}{\bibfnamefont{S.~J.} \bibnamefont{Brodsky}},
  \bibinfo{journal}{Phys. Lett.} \textbf{\bibinfo{volume}{B87}},
  \bibinfo{pages}{359}
 (\bibinfo{year}{1979}).

\bibitem[{\citenamefont{Lepage and Brodsky}(1980)}]{Lepage:1980fj}
\bibinfo{author}{\bibfnamefont{G.~P.} \bibnamefont{Lepage}} \bibnamefont{and}
  \bibinfo{author}{\bibfnamefont{S.~J.} \bibnamefont{Brodsky}},
  \bibinfo{journal}{Phys. Rev.} \textbf{\bibinfo{volume}{D22}},
  \bibinfo{pages}{2157}
 (\bibinfo{year}{1980}).

\bibitem[{\citenamefont{Chernyak et~al.}(1977)\citenamefont{Chernyak,
  Zhitnitsky, and Serbo}}]{Chernyak:1977fk}
\bibinfo{author}{\bibfnamefont{V.~L.} \bibnamefont{Chernyak}},
  \bibinfo{author}{\bibfnamefont{A.~R.} \bibnamefont{Zhitnitsky}},
  \bibnamefont{and} \bibinfo{author}{\bibfnamefont{V.~G.} \bibnamefont{Serbo}},
  \bibinfo{journal}{JETP Lett.} \textbf{\bibinfo{volume}{26}},
  \bibinfo{pages}{594}
 (\bibinfo{year}{1977}).

\bibitem[{\citenamefont{Chernyak et~al.}(1980)\citenamefont{Chernyak, Serbo,
  and Zhitnitsky}}]{Chernyak:1980dk}
\bibinfo{author}{\bibfnamefont{V.~L.} \bibnamefont{Chernyak}},
  \bibinfo{author}{\bibfnamefont{V.~G.} \bibnamefont{Serbo}}, \bibnamefont{and}
  \bibinfo{author}{\bibfnamefont{A.~R.} \bibnamefont{Zhitnitsky}},
  \bibinfo{journal}{Sov. J. Nucl. Phys.} \textbf{\bibinfo{volume}{31}},
  \bibinfo{pages}{552}
 (\bibinfo{year}{1980}).

\bibitem[{\citenamefont{Stefanis et~al.}(2000)\citenamefont{Stefanis, Schroers,
  and Kim}}]{Stefanis:2000vd}
\bibinfo{author}{\bibfnamefont{N.~G.} \bibnamefont{Stefanis}},
  \bibinfo{author}{\bibfnamefont{W.}~\bibnamefont{Schroers}}, \bibnamefont{and}
  \bibinfo{author}{\bibfnamefont{H.-C.} \bibnamefont{Kim}},
  \bibinfo{journal}{Eur. Phys. J.} \textbf{\bibinfo{volume}{C18}},
  \bibinfo{pages}{137} (\bibinfo{year}{2000}),
\eprint{hep-ph/0005218}.

\bibitem[{\citenamefont{Beneke et~al.}(2000)\citenamefont{Beneke, Buchalla,
  Neubert, and Sachrajda}}]{Beneke:2000ry}
\bibinfo{author}{\bibfnamefont{M.}~\bibnamefont{Beneke}},
  \bibinfo{author}{\bibfnamefont{G.}~\bibnamefont{Buchalla}},
  \bibinfo{author}{\bibfnamefont{M.}~\bibnamefont{Neubert}}, \bibnamefont{and}
  \bibinfo{author}{\bibfnamefont{C.~T.} \bibnamefont{Sachrajda}},
  \bibinfo{journal}{Nucl. Phys.} \textbf{\bibinfo{volume}{B591}},
  \bibinfo{pages}{313} (\bibinfo{year}{2000}),
\eprint{hep-ph/0006124}.

\bibitem[{\citenamefont{Keum and Li}(2001)}]{Keum:2000ms}
\bibinfo{author}{\bibfnamefont{Y.-Y.} \bibnamefont{Keum}} \bibnamefont{and}
  \bibinfo{author}{\bibfnamefont{H.-n.} \bibnamefont{Li}},
  \bibinfo{journal}{Phys. Rev.} \textbf{\bibinfo{volume}{D63}},
  \bibinfo{pages}{074006} (\bibinfo{year}{2001}),
\eprint{hep-ph/0006001}.

\bibitem[{\citenamefont{Bauer et~al.}(2001)\citenamefont{Bauer, Fleming,
  Pirjol, and Stewart}}]{Bauer:2000yr}
\bibinfo{author}{\bibfnamefont{C.~W.} \bibnamefont{Bauer}},
  \bibinfo{author}{\bibfnamefont{S.}~\bibnamefont{Fleming}},
  \bibinfo{author}{\bibfnamefont{D.}~\bibnamefont{Pirjol}}, \bibnamefont{and}
  \bibinfo{author}{\bibfnamefont{I.~W.} \bibnamefont{Stewart}},
  \bibinfo{journal}{Phys. Rev.} \textbf{\bibinfo{volume}{D63}},
  \bibinfo{pages}{114020} (\bibinfo{year}{2001}),
\eprint{hep-ph/0011336}.

\bibitem[{\citenamefont{Bauer et~al.}(2002)\citenamefont{Bauer, Pirjol, and
  Stewart}}]{Bauer:2001yt}
\bibinfo{author}{\bibfnamefont{C.~W.} \bibnamefont{Bauer}},
  \bibinfo{author}{\bibfnamefont{D.}~\bibnamefont{Pirjol}}, \bibnamefont{and}
  \bibinfo{author}{\bibfnamefont{I.~W.} \bibnamefont{Stewart}},
  \bibinfo{journal}{Phys. Rev.} \textbf{\bibinfo{volume}{D65}},
  \bibinfo{pages}{054022} (\bibinfo{year}{2002}),
\eprint{hep-ph/0109045}.

\bibitem[{\citenamefont{Ball and Braun}(1998)}]{Ball:1998kk}
\bibinfo{author}{\bibfnamefont{P.}~\bibnamefont{Ball}} \bibnamefont{and}
  \bibinfo{author}{\bibfnamefont{V.~M.} \bibnamefont{Braun}},
  \bibinfo{journal}{Phys. Rev.} \textbf{\bibinfo{volume}{D58}},
  \bibinfo{pages}{094016} (\bibinfo{year}{1998}),
\eprint{hep-ph/9805422}.

\bibitem[{\citenamefont{Khodjamirian et~al.}(2000)\citenamefont{Khodjamirian,
  Ruckl, Weinzierl, Winhart, and Yakovlev}}]{Khodjamirian:2000ds}
\bibinfo{author}{\bibfnamefont{A.}~\bibnamefont{Khodjamirian}},
  \bibinfo{author}{\bibfnamefont{R.}~\bibnamefont{Ruckl}},
  \bibinfo{author}{\bibfnamefont{S.}~\bibnamefont{Weinzierl}},
  \bibinfo{author}{\bibfnamefont{C.~W.} \bibnamefont{Winhart}},
  \bibnamefont{and} \bibinfo{author}{\bibfnamefont{O.~I.}
  \bibnamefont{Yakovlev}}, \bibinfo{journal}{Phys. Rev.}
  \textbf{\bibinfo{volume}{D62}}, \bibinfo{pages}{114002}
  (\bibinfo{year}{2000}),
\eprint{hep-ph/0001297}.

\bibitem[{\citenamefont{Ball and Zwicky}(2005{\natexlab{a}})}]{Ball:2004rg}
\bibinfo{author}{\bibfnamefont{P.}~\bibnamefont{Ball}} \bibnamefont{and}
  \bibinfo{author}{\bibfnamefont{R.}~\bibnamefont{Zwicky}},
  \bibinfo{journal}{Phys. Rev.} \textbf{\bibinfo{volume}{D71}},
  \bibinfo{pages}{014029} (\bibinfo{year}{2005}{\natexlab{a}}),
\eprint{hep-ph/0412079}.

\bibitem[{\citenamefont{Eidelman et~al.}(2004)}]{Eidelman:2004wy}
\bibinfo{author}{\bibfnamefont{S.}~\bibnamefont{Eidelman}} \bibnamefont{et~al.}
  (\bibinfo{collaboration}{Particle Data Group}), \bibinfo{journal}{Phys.
  Lett.} \textbf{\bibinfo{volume}{B592}}, \bibinfo{pages}{1}
 (\bibinfo{year}{2004}).

\bibitem[{\citenamefont{Brodsky et~al.}(1980)\citenamefont{Brodsky, Frishman,
  Lepage, and Sachrajda}}]{Brodsky:1980ny}
\bibinfo{author}{\bibfnamefont{S.~J.} \bibnamefont{Brodsky}},
  \bibinfo{author}{\bibfnamefont{Y.}~\bibnamefont{Frishman}},
  \bibinfo{author}{\bibfnamefont{G.~P.} \bibnamefont{Lepage}},
  \bibnamefont{and} \bibinfo{author}{\bibfnamefont{C.~T.}
  \bibnamefont{Sachrajda}}, \bibinfo{journal}{Phys. Lett.}
  \textbf{\bibinfo{volume}{B91}}, \bibinfo{pages}{239}
 (\bibinfo{year}{1980}).

\bibitem[{\citenamefont{Ohrndorf}(1982)}]{Ohrndorf:1981qv}
\bibinfo{author}{\bibfnamefont{T.}~\bibnamefont{Ohrndorf}},
  \bibinfo{journal}{Nucl. Phys.} \textbf{\bibinfo{volume}{B198}},
  \bibinfo{pages}{26}
 (\bibinfo{year}{1982}).

\bibitem[{\citenamefont{Braun et~al.}(2003)\citenamefont{Braun, Korchemsky, and
  M{\"u}ller}}]{Braun:2003rp}
\bibinfo{author}{\bibfnamefont{V.~M.} \bibnamefont{Braun}},
  \bibinfo{author}{\bibfnamefont{G.~P.} \bibnamefont{Korchemsky}},
  \bibnamefont{and}
  \bibinfo{author}{\bibfnamefont{D.}~\bibnamefont{M{\"u}ller}},
  \bibinfo{journal}{Prog. Part. Nucl. Phys.} \textbf{\bibinfo{volume}{51}},
  \bibinfo{pages}{311} (\bibinfo{year}{2003}),
\eprint{hep-ph/0306057}.

\bibitem[{\citenamefont{Makeenko}(1981)}]{Makeenko:1980bh}
\bibinfo{author}{\bibfnamefont{Y.~M.} \bibnamefont{Makeenko}},
  \bibinfo{journal}{Sov. J. Nucl. Phys.} \textbf{\bibinfo{volume}{33}},
  \bibinfo{pages}{440}
 (\bibinfo{year}{1981}).

\bibitem[{\citenamefont{Mikhailov and Radyushkin}(1985)}]{Mikhailov:1984ii}
\bibinfo{author}{\bibfnamefont{S.~V.} \bibnamefont{Mikhailov}}
  \bibnamefont{and} \bibinfo{author}{\bibfnamefont{A.~V.}
  \bibnamefont{Radyushkin}}, \bibinfo{journal}{Nucl. Phys.}
  \textbf{\bibinfo{volume}{B254}}, \bibinfo{pages}{89}
 (\bibinfo{year}{1985}).

\bibitem[{\citenamefont{Mueller}(1994)}]{Mueller:1993hg}
\bibinfo{author}{\bibfnamefont{D.}~\bibnamefont{Mueller}},
  \bibinfo{journal}{Phys. Rev.} \textbf{\bibinfo{volume}{D49}},
  \bibinfo{pages}{2525}
 (\bibinfo{year}{1994}).

\bibitem[{\citenamefont{Mueller}(1995)}]{Mueller:1994cn}
\bibinfo{author}{\bibfnamefont{D.}~\bibnamefont{Mueller}},
  \bibinfo{journal}{Phys. Rev.} \textbf{\bibinfo{volume}{D51}},
  \bibinfo{pages}{3855} (\bibinfo{year}{1995}),
\eprint{hep-ph/9411338}.

\bibitem[{\citenamefont{Mertig and van Neerven}(1996)}]{Mertig:1995ny}
\bibinfo{author}{\bibfnamefont{R.}~\bibnamefont{Mertig}} \bibnamefont{and}
  \bibinfo{author}{\bibfnamefont{W.~L.} \bibnamefont{van Neerven}},
  \bibinfo{journal}{Z. Phys.} \textbf{\bibinfo{volume}{C70}},
  \bibinfo{pages}{637} (\bibinfo{year}{1996}),
\eprint{hep-ph/9506451}.

\bibitem[{\citenamefont{Bakulev et~al.}(2004)\citenamefont{Bakulev,
  Passek-Kumericki, Schroers, and Stefanis}}]{Bakulev:2004cu}
\bibinfo{author}{\bibfnamefont{A.~P.} \bibnamefont{Bakulev}},
  \bibinfo{author}{\bibfnamefont{K.}~\bibnamefont{Passek-Kumericki}},
  \bibinfo{author}{\bibfnamefont{W.}~\bibnamefont{Schroers}}, \bibnamefont{and}
  \bibinfo{author}{\bibfnamefont{N.~G.} \bibnamefont{Stefanis}},
  \bibinfo{journal}{Phys. Rev.} \textbf{\bibinfo{volume}{D70}},
  \bibinfo{pages}{033014} (\bibinfo{year}{2004}),
\eprint{hep-ph/0405062}.

\bibitem[{\citenamefont{Chernyak and Zhitnitsky}(1982)}]{Chernyak:1981zz}
\bibinfo{author}{\bibfnamefont{V.~L.} \bibnamefont{Chernyak}} \bibnamefont{and}
  \bibinfo{author}{\bibfnamefont{A.~R.} \bibnamefont{Zhitnitsky}},
  \bibinfo{journal}{Nucl. Phys.} \textbf{\bibinfo{volume}{B201}},
  \bibinfo{pages}{492}
 (\bibinfo{year}{1982}).

\bibitem[{\citenamefont{Chernyak and Zhitnitsky}(1984)}]{Chernyak:1983ej}
\bibinfo{author}{\bibfnamefont{V.~L.} \bibnamefont{Chernyak}} \bibnamefont{and}
  \bibinfo{author}{\bibfnamefont{A.~R.} \bibnamefont{Zhitnitsky}},
  \bibinfo{journal}{Phys. Rept.} \textbf{\bibinfo{volume}{112}},
  \bibinfo{pages}{173}
 (\bibinfo{year}{1984}).

\bibitem[{\citenamefont{Stefanis et~al.}(1999)\citenamefont{Stefanis, Schroers,
  and Kim}}]{Stefanis:1998dg}
\bibinfo{author}{\bibfnamefont{N.~G.} \bibnamefont{Stefanis}},
  \bibinfo{author}{\bibfnamefont{W.}~\bibnamefont{Schroers}}, \bibnamefont{and}
  \bibinfo{author}{\bibfnamefont{H.-C.} \bibnamefont{Kim}},
  \bibinfo{journal}{Phys. Lett.} \textbf{\bibinfo{volume}{B449}},
  \bibinfo{pages}{299} (\bibinfo{year}{1999}),
\eprint{hep-ph/9807298}.

\bibitem[{\citenamefont{Ball et~al.}(2006)\citenamefont{Ball, Braun, and
  Lenz}}]{Ball:2006wn}
\bibinfo{author}{\bibfnamefont{P.}~\bibnamefont{Ball}},
  \bibinfo{author}{\bibfnamefont{V.~M.} \bibnamefont{Braun}}, \bibnamefont{and}
  \bibinfo{author}{\bibfnamefont{A.}~\bibnamefont{Lenz}}
  (\bibinfo{year}{2006}),
\eprint{hep-ph/0603063}.

\bibitem[{\citenamefont{Bakulev et~al.}(2003)\citenamefont{Bakulev, Mikhailov,
  and Stefanis}}]{Bakulev:2002uc}
\bibinfo{author}{\bibfnamefont{A.~P.} \bibnamefont{Bakulev}},
  \bibinfo{author}{\bibfnamefont{S.~V.} \bibnamefont{Mikhailov}},
  \bibnamefont{and} \bibinfo{author}{\bibfnamefont{N.~G.}
  \bibnamefont{Stefanis}}, \bibinfo{journal}{Phys. Rev.}
  \textbf{\bibinfo{volume}{D67}}, \bibinfo{pages}{074012}
  (\bibinfo{year}{2003}),
\eprint{hep-ph/0212250}.

\bibitem[{\citenamefont{Khodjamirian et~al.}(2004)\citenamefont{Khodjamirian,
  Mannel, and Melcher}}]{Khodjamirian:2004ga}
\bibinfo{author}{\bibfnamefont{A.}~\bibnamefont{Khodjamirian}},
  \bibinfo{author}{\bibfnamefont{T.}~\bibnamefont{Mannel}}, \bibnamefont{and}
  \bibinfo{author}{\bibfnamefont{M.}~\bibnamefont{Melcher}},
  \bibinfo{journal}{Phys. Rev.} \textbf{\bibinfo{volume}{D70}},
  \bibinfo{pages}{094002} (\bibinfo{year}{2004}),
\eprint{hep-ph/0407226}.

\bibitem[{\citenamefont{Braun and Lenz}(2004)}]{Braun:2004vf}
\bibinfo{author}{\bibfnamefont{V.~M.} \bibnamefont{Braun}} \bibnamefont{and}
  \bibinfo{author}{\bibfnamefont{A.}~\bibnamefont{Lenz}},
  \bibinfo{journal}{Phys. Rev.} \textbf{\bibinfo{volume}{D70}},
  \bibinfo{pages}{074020} (\bibinfo{year}{2004}),
\eprint{hep-ph/0407282}.

\bibitem[{\citenamefont{Ball and Zwicky}(2006{\natexlab{a}})}]{Ball:2005vx}
\bibinfo{author}{\bibfnamefont{P.}~\bibnamefont{Ball}} \bibnamefont{and}
  \bibinfo{author}{\bibfnamefont{R.}~\bibnamefont{Zwicky}},
  \bibinfo{journal}{Phys. Lett.} \textbf{\bibinfo{volume}{B633}},
  \bibinfo{pages}{289} (\bibinfo{year}{2006}{\natexlab{a}}),
\eprint{hep-ph/0510338}.

\bibitem[{\citenamefont{Ball and Zwicky}(2006{\natexlab{b}})}]{Ball:2006fz}
\bibinfo{author}{\bibfnamefont{P.}~\bibnamefont{Ball}} \bibnamefont{and}
  \bibinfo{author}{\bibfnamefont{R.}~\bibnamefont{Zwicky}},
  \bibinfo{journal}{JHEP} \textbf{\bibinfo{volume}{02}}, \bibinfo{pages}{034}
  (\bibinfo{year}{2006}{\natexlab{b}}),
\eprint{hep-ph/0601086}.

\bibitem[{\citenamefont{Schmedding and Yakovlev}(2000)}]{Schmedding:1999ap}
\bibinfo{author}{\bibfnamefont{A.}~\bibnamefont{Schmedding}} \bibnamefont{and}
  \bibinfo{author}{\bibfnamefont{O.~I.} \bibnamefont{Yakovlev}},
  \bibinfo{journal}{Phys. Rev.} \textbf{\bibinfo{volume}{D62}},
  \bibinfo{pages}{116002} (\bibinfo{year}{2000}),
\eprint{hep-ph/9905392}.

\bibitem[{\citenamefont{Bakulev et~al.}(2006)\citenamefont{Bakulev, Mikhailov,
  and Stefanis}}]{Bakulev:2005cp}
\bibinfo{author}{\bibfnamefont{A.~P.} \bibnamefont{Bakulev}},
  \bibinfo{author}{\bibfnamefont{S.~V.} \bibnamefont{Mikhailov}},
  \bibnamefont{and} \bibinfo{author}{\bibfnamefont{N.~G.}
  \bibnamefont{Stefanis}}, \bibinfo{journal}{Phys. Rev.}
  \textbf{\bibinfo{volume}{D73}}, \bibinfo{pages}{056002}
  (\bibinfo{year}{2006}),
\eprint{hep-ph/0512119}.

\bibitem[{\citenamefont{Agaev}(2005)}]{Agaev:2005rc}
\bibinfo{author}{\bibfnamefont{S.~S.} \bibnamefont{Agaev}},
  \bibinfo{journal}{Phys. Rev.} \textbf{\bibinfo{volume}{D72}},
  \bibinfo{pages}{114010} (\bibinfo{year}{2005}),
\eprint{hep-ph/0511192}.

\bibitem[{\citenamefont{Ball and Zwicky}(2005{\natexlab{b}})}]{Ball:2005tb}
\bibinfo{author}{\bibfnamefont{P.}~\bibnamefont{Ball}} \bibnamefont{and}
  \bibinfo{author}{\bibfnamefont{R.}~\bibnamefont{Zwicky}},
  \bibinfo{journal}{Phys. Lett.} \textbf{\bibinfo{volume}{B625}},
  \bibinfo{pages}{225} (\bibinfo{year}{2005}{\natexlab{b}}),
\eprint{hep-ph/0507076}.

\bibitem[{\citenamefont{Bakulev et~al.}(2001)\citenamefont{Bakulev, Mikhailov,
  and Stefanis}}]{Bakulev:2001pa}
\bibinfo{author}{\bibfnamefont{A.~P.} \bibnamefont{Bakulev}},
  \bibinfo{author}{\bibfnamefont{S.~V.} \bibnamefont{Mikhailov}},
  \bibnamefont{and} \bibinfo{author}{\bibfnamefont{N.~G.}
  \bibnamefont{Stefanis}}, \bibinfo{journal}{Phys. Lett.}
  \textbf{\bibinfo{volume}{B508}}, \bibinfo{pages}{279} (\bibinfo{year}{2001}),
\eprint{hep-ph/0103119}.

\bibitem[{\citenamefont{Dalley and van~de Sande}(2003)}]{Dalley:2002nj}
\bibinfo{author}{\bibfnamefont{S.}~\bibnamefont{Dalley}} \bibnamefont{and}
  \bibinfo{author}{\bibfnamefont{B.}~\bibnamefont{van~de Sande}},
  \bibinfo{journal}{Phys. Rev.} \textbf{\bibinfo{volume}{D67}},
  \bibinfo{pages}{114507} (\bibinfo{year}{2003}),
\eprint{hep-ph/0212086}.

\bibitem[{\citenamefont{Capitani et~al.}(1999)}]{Capitani:1999fm}
\bibinfo{author}{\bibfnamefont{S.}~\bibnamefont{Capitani}}
  \bibnamefont{et~al.}, \bibinfo{journal}{Nucl. Phys. Proc. Suppl.}
  \textbf{\bibinfo{volume}{79}}, \bibinfo{pages}{173} (\bibinfo{year}{1999}),
\eprint{hep-ph/9906320}.

\bibitem[{\citenamefont{Detmold and Lin}(2006)}]{Detmold:2005gg}
\bibinfo{author}{\bibfnamefont{W.}~\bibnamefont{Detmold}} \bibnamefont{and}
  \bibinfo{author}{\bibfnamefont{C.~J.~D.} \bibnamefont{Lin}},
  \bibinfo{journal}{Phys. Rev.} \textbf{\bibinfo{volume}{D73}},
  \bibinfo{pages}{014501} (\bibinfo{year}{2006}),
\eprint{hep-lat/0507007}.

\bibitem[{\citenamefont{Braun and Filyanov}(1989)}]{Braun:1988qv}
\bibinfo{author}{\bibfnamefont{V.~M.} \bibnamefont{Braun}} \bibnamefont{and}
  \bibinfo{author}{\bibfnamefont{I.~E.} \bibnamefont{Filyanov}},
  \bibinfo{journal}{Z. Phys.} \textbf{\bibinfo{volume}{C44}},
  \bibinfo{pages}{157}
 (\bibinfo{year}{1989}).

\bibitem[{\citenamefont{Aitala et~al.}(2001)}]{Aitala:2000hb}
\bibinfo{author}{\bibfnamefont{E.~M.} \bibnamefont{Aitala}}
  \bibnamefont{et~al.} (\bibinfo{collaboration}{E791}), \bibinfo{journal}{Phys.
  Rev. Lett.} \textbf{\bibinfo{volume}{86}}, \bibinfo{pages}{4768}
  (\bibinfo{year}{2001}),
\eprint{hep-ex/0010043}.

\bibitem[{\citenamefont{Braun et~al.}(2001)\citenamefont{Braun, Ivanov,
  Sch{\"a}fer, and Szymanowski}}]{Braun:2001ih}
\bibinfo{author}{\bibfnamefont{V.~M.} \bibnamefont{Braun}},
  \bibinfo{author}{\bibfnamefont{D.~Y.} \bibnamefont{Ivanov}},
  \bibinfo{author}{\bibfnamefont{A.}~\bibnamefont{Sch{\"a}fer}},
  \bibnamefont{and}
  \bibinfo{author}{\bibfnamefont{L.}~\bibnamefont{Szymanowski}},
  \bibinfo{journal}{Phys. Lett.} \textbf{\bibinfo{volume}{B509}},
  \bibinfo{pages}{43} (\bibinfo{year}{2001}),
\eprint{hep-ph/0103275}.

\bibitem[{\citenamefont{Braun et~al.}(2002)\citenamefont{Braun, Ivanov,
  Sch{\"a}fer, and Szymanowski}}]{Braun:2002wu}
\bibinfo{author}{\bibfnamefont{V.~M.} \bibnamefont{Braun}},
  \bibinfo{author}{\bibfnamefont{D.~Y.} \bibnamefont{Ivanov}},
  \bibinfo{author}{\bibfnamefont{A.}~\bibnamefont{Sch{\"a}fer}},
  \bibnamefont{and}
  \bibinfo{author}{\bibfnamefont{L.}~\bibnamefont{Szymanowski}},
  \bibinfo{journal}{Nucl. Phys.} \textbf{\bibinfo{volume}{B638}},
  \bibinfo{pages}{111} (\bibinfo{year}{2002}),
\eprint{hep-ph/0204191}.

\bibitem[{\citenamefont{Chernyak}(2001)}]{Chernyak:2001ph}
\bibinfo{author}{\bibfnamefont{V.}~\bibnamefont{Chernyak}},
  \bibinfo{journal}{Phys. Lett.} \textbf{\bibinfo{volume}{B516}},
  \bibinfo{pages}{116} (\bibinfo{year}{2001}),
\eprint{hep-ph/0103295}.

\bibitem[{\citenamefont{Chernyak and Grozin}(2001)}]{Chernyak:2001wk}
\bibinfo{author}{\bibfnamefont{V.~L.} \bibnamefont{Chernyak}} \bibnamefont{and}
  \bibinfo{author}{\bibfnamefont{A.~G.} \bibnamefont{Grozin}},
  \bibinfo{journal}{Phys. Lett.} \textbf{\bibinfo{volume}{B517}},
  \bibinfo{pages}{119} (\bibinfo{year}{2001}),
\eprint{hep-ph/0106162}.

\bibitem[{\citenamefont{Kronfeld and Photiadis}(1985)}]{Kronfeld:1984zv}
\bibinfo{author}{\bibfnamefont{A.~S.} \bibnamefont{Kronfeld}} \bibnamefont{and}
  \bibinfo{author}{\bibfnamefont{D.~M.} \bibnamefont{Photiadis}},
  \bibinfo{journal}{Phys. Rev.} \textbf{\bibinfo{volume}{D31}},
  \bibinfo{pages}{2939}
 (\bibinfo{year}{1985}).

\bibitem[{\citenamefont{Martinelli and Sachrajda}(1987)}]{Martinelli:1987si}
\bibinfo{author}{\bibfnamefont{G.}~\bibnamefont{Martinelli}} \bibnamefont{and}
  \bibinfo{author}{\bibfnamefont{C.~T.} \bibnamefont{Sachrajda}},
  \bibinfo{journal}{Phys. Lett.} \textbf{\bibinfo{volume}{B190}},
  \bibinfo{pages}{151}
 (\bibinfo{year}{1987}).

\bibitem[{\citenamefont{Daniel et~al.}(1991)\citenamefont{Daniel, Gupta, and
  Richards}}]{Daniel:1990ah}
\bibinfo{author}{\bibfnamefont{D.}~\bibnamefont{Daniel}},
  \bibinfo{author}{\bibfnamefont{R.}~\bibnamefont{Gupta}}, \bibnamefont{and}
  \bibinfo{author}{\bibfnamefont{D.~G.} \bibnamefont{Richards}},
  \bibinfo{journal}{Phys. Rev.} \textbf{\bibinfo{volume}{D43}},
  \bibinfo{pages}{3715}
 (\bibinfo{year}{1991}).

\bibitem[{\citenamefont{Del~Debbio et~al.}(2000)\citenamefont{Del~Debbio,
  Di~Pierro, Dougall, and Sachrajda}}]{DelDebbio:1999mq}
\bibinfo{author}{\bibfnamefont{L.}~\bibnamefont{Del~Debbio}},
  \bibinfo{author}{\bibfnamefont{M.}~\bibnamefont{Di~Pierro}},
  \bibinfo{author}{\bibfnamefont{A.}~\bibnamefont{Dougall}}, \bibnamefont{and}
  \bibinfo{author}{\bibfnamefont{C.~T.} \bibnamefont{Sachrajda}}
  (\bibinfo{collaboration}{UKQCD}), \bibinfo{journal}{Nucl. Phys. Proc. Suppl.}
  \textbf{\bibinfo{volume}{83}}, \bibinfo{pages}{235} (\bibinfo{year}{2000}),
\eprint{hep-lat/9909147}.

\bibitem[{\citenamefont{Del~Debbio et~al.}(2003)\citenamefont{Del~Debbio,
  Di~Pierro, and Dougall}}]{DelDebbio:2002mq}
\bibinfo{author}{\bibfnamefont{L.}~\bibnamefont{Del~Debbio}},
  \bibinfo{author}{\bibfnamefont{M.}~\bibnamefont{Di~Pierro}},
  \bibnamefont{and} \bibinfo{author}{\bibfnamefont{A.}~\bibnamefont{Dougall}},
  \bibinfo{journal}{Nucl. Phys. Proc. Suppl.} \textbf{\bibinfo{volume}{119}},
  \bibinfo{pages}{416} (\bibinfo{year}{2003}),
\eprint{hep-lat/0211037}.

\bibitem[{\citenamefont{G{\"o}ckeler
  et~al.}(2005{\natexlab{a}})}]{Gockeler:2005jz}
\bibinfo{author}{\bibfnamefont{M.}~\bibnamefont{G{\"o}ckeler}}
  \bibnamefont{et~al.} (\bibinfo{year}{2005}{\natexlab{a}}),
\eprint{hep-lat/0510089}.

\bibitem[{\citenamefont{Best et~al.}(1997)}]{Best:1997qp}
\bibinfo{author}{\bibfnamefont{C.}~\bibnamefont{Best}} \bibnamefont{et~al.},
  \bibinfo{journal}{Phys. Rev.} \textbf{\bibinfo{volume}{D56}},
  \bibinfo{pages}{2743} (\bibinfo{year}{1997}),
\eprint{hep-lat/9703014}.

\bibitem[{\citenamefont{G{\"o}ckeler
  et~al.}(2006{\natexlab{a}})}]{Gockeler:2006nb}
\bibinfo{author}{\bibfnamefont{M.}~\bibnamefont{G{\"o}ckeler}}
  \bibnamefont{et~al.} (\bibinfo{year}{2006}{\natexlab{a}}),
\eprint{hep-lat/0605002}.

\bibitem[{\citenamefont{Capitani et~al.}(2001)}]{Capitani:2000xi}
\bibinfo{author}{\bibfnamefont{S.}~\bibnamefont{Capitani}}
  \bibnamefont{et~al.}, \bibinfo{journal}{Nucl. Phys.}
  \textbf{\bibinfo{volume}{B593}}, \bibinfo{pages}{183} (\bibinfo{year}{2001}),
\eprint{hep-lat/0007004}.

\bibitem[{\citenamefont{G{\"o}ckeler
  et~al.}(2005{\natexlab{b}})}]{Gockeler:2004xb}
\bibinfo{author}{\bibfnamefont{M.}~\bibnamefont{G{\"o}ckeler}}
  \bibnamefont{et~al.}, \bibinfo{journal}{Nucl. Phys.}
  \textbf{\bibinfo{volume}{B717}}, \bibinfo{pages}{304}
  (\bibinfo{year}{2005}{\natexlab{b}}),
\eprint{hep-lat/0410009}.

\bibitem[{\citenamefont{Aubin et~al.}(2004)}]{Aubin:2004wf}
\bibinfo{author}{\bibfnamefont{C.}~\bibnamefont{Aubin}} \bibnamefont{et~al.},
  \bibinfo{journal}{Phys. Rev.} \textbf{\bibinfo{volume}{D70}},
  \bibinfo{pages}{094505} (\bibinfo{year}{2004}),
\eprint{hep-lat/0402030}.

\bibitem[{\citenamefont{Khan et~al.}(2006)}]{Khan:2006de}
\bibinfo{author}{\bibfnamefont{A.~A.} \bibnamefont{Khan}} \bibnamefont{et~al.}
  (\bibinfo{year}{2006}),
\eprint{hep-lat/0603028}.

\bibitem[{\citenamefont{G{\"o}ckeler
  et~al.}(2006{\natexlab{b}})}]{Gockeler:2005rv}
\bibinfo{author}{\bibfnamefont{M.}~\bibnamefont{G{\"o}ckeler}}
  \bibnamefont{et~al.}, \bibinfo{journal}{Phys. Rev.}
  \textbf{\bibinfo{volume}{D73}}, \bibinfo{pages}{014513}
  (\bibinfo{year}{2006}{\natexlab{b}}),
\eprint{hep-ph/0502212}.

\bibitem[{\citenamefont{G{\"o}ckeler
  et~al.}(2005{\natexlab{c}})\citenamefont{G{\"o}ckeler, Horsley, Pleiter,
  Rakow, and Schierholz}}]{Gockeler:2004wp}
\bibinfo{author}{\bibfnamefont{M.}~\bibnamefont{G{\"o}ckeler}},
  \bibinfo{author}{\bibfnamefont{R.}~\bibnamefont{Horsley}},
  \bibinfo{author}{\bibfnamefont{D.}~\bibnamefont{Pleiter}},
  \bibinfo{author}{\bibfnamefont{P.~E.~L.} \bibnamefont{Rakow}},
  \bibnamefont{and}
  \bibinfo{author}{\bibfnamefont{G.}~\bibnamefont{Schierholz}}
  (\bibinfo{collaboration}{QCDSF}), \bibinfo{journal}{Phys. Rev.}
  \textbf{\bibinfo{volume}{D71}}, \bibinfo{pages}{114511}
  (\bibinfo{year}{2005}{\natexlab{c}}),
\eprint{hep-ph/0410187}.

\bibitem[{\citenamefont{Martinelli et~al.}(1995)\citenamefont{Martinelli,
  Pittori, Sachrajda, Testa, and Vladikas}}]{Martinelli:1994ty}
\bibinfo{author}{\bibfnamefont{G.}~\bibnamefont{Martinelli}},
  \bibinfo{author}{\bibfnamefont{C.}~\bibnamefont{Pittori}},
  \bibinfo{author}{\bibfnamefont{C.~T.} \bibnamefont{Sachrajda}},
  \bibinfo{author}{\bibfnamefont{M.}~\bibnamefont{Testa}}, \bibnamefont{and}
  \bibinfo{author}{\bibfnamefont{A.}~\bibnamefont{Vladikas}},
  \bibinfo{journal}{Nucl. Phys.} \textbf{\bibinfo{volume}{B445}},
  \bibinfo{pages}{81} (\bibinfo{year}{1995}),
\eprint{hep-lat/9411010}.

\bibitem[{\citenamefont{G{\"o}ckeler et~al.}(1999)}]{Gockeler:1998ye}
\bibinfo{author}{\bibfnamefont{M.}~\bibnamefont{G{\"o}ckeler}}
  \bibnamefont{et~al.}, \bibinfo{journal}{Nucl. Phys.}
  \textbf{\bibinfo{volume}{B544}}, \bibinfo{pages}{699} (\bibinfo{year}{1999}),
\eprint{hep-lat/9807044}.

\bibitem[{\citenamefont{G{\"o}ckeler
  et~al.}(2005{\natexlab{d}})}]{Gockeler:2005vw}
\bibinfo{author}{\bibfnamefont{M.}~\bibnamefont{G{\"o}ckeler}}
  \bibnamefont{et~al.}, \bibinfo{journal}{Phys. Rev.}
  \textbf{\bibinfo{volume}{D72}}, \bibinfo{pages}{054507}
  (\bibinfo{year}{2005}{\natexlab{d}}),
\eprint{hep-lat/0506017}.

\bibitem[{\citenamefont{Chen and Stewart}(2004)}]{Chen:2003fp}
\bibinfo{author}{\bibfnamefont{J.-W.} \bibnamefont{Chen}} \bibnamefont{and}
  \bibinfo{author}{\bibfnamefont{I.~W.} \bibnamefont{Stewart}},
  \bibinfo{journal}{Phys. Rev. Lett.} \textbf{\bibinfo{volume}{92}},
  \bibinfo{pages}{202001} (\bibinfo{year}{2004}),
\eprint{hep-ph/0311285}.

\bibitem[{\citenamefont{Chen et~al.}(2006)\citenamefont{Chen, Tsai, and
  Weng}}]{Chen:2005js}
\bibinfo{author}{\bibfnamefont{J.-W.} \bibnamefont{Chen}},
  \bibinfo{author}{\bibfnamefont{H.-M.} \bibnamefont{Tsai}}, \bibnamefont{and}
  \bibinfo{author}{\bibfnamefont{K.-C.} \bibnamefont{Weng}},
  \bibinfo{journal}{Phys. Rev.} \textbf{\bibinfo{volume}{D73}},
  \bibinfo{pages}{054010} (\bibinfo{year}{2006}),
\eprint{hep-ph/0511036}.

\bibitem[{\citenamefont{Boyle et~al.}(2006)}]{Boyle:2006pw}
\bibinfo{author}{\bibfnamefont{P.~A.} \bibnamefont{Boyle}} \bibnamefont{et~al.}
  (\bibinfo{year}{2006}),
\eprint{hep-lat/0607018}.

\bibitem[{\citenamefont{Allton et~al.}(2002)}]{Allton:2001sk}
\bibinfo{author}{\bibfnamefont{C.~R.} \bibnamefont{Allton}}
  \bibnamefont{et~al.} (\bibinfo{collaboration}{UKQCD}),
  \bibinfo{journal}{Phys. Rev.} \textbf{\bibinfo{volume}{D65}},
  \bibinfo{pages}{054502} (\bibinfo{year}{2002}),
\eprint{hep-lat/0107021}.

\end{thebibliography}


\end{document}